\documentclass[acmtog]{acmart}
\setcitestyle{square}

\usepackage{booktabs} 
\usepackage{bm}
\usepackage{amsfonts}
\usepackage{amssymb}
\usepackage{algorithmic}
\usepackage{amsmath}
\usepackage{graphicx}
\usepackage{caption}
\usepackage{subfigure}
\usepackage{array}
\usepackage{enumitem}
\usepackage{multirow}
\usepackage{xspace}
\usepackage{gensymb}

\usepackage[ruled]{algorithm2e} 

\SetAlFnt{\small}
\SetAlCapFnt{\small}
\SetAlCapNameFnt{\small}
\SetAlCapHSkip{0pt}
\IncMargin{-\parindent}

\title[DeepMimic: Example-Guided Deep Reinforcement Learning of Physics-Based Character Skills]{DeepMimic: Example-Guided Deep Reinforcement Learning\\ of Physics-Based Character Skills}

\author{Xue Bin Peng}
\affiliation{\institution{University of California, Berkeley}}
\author{Pieter Abbeel}
\affiliation{\institution{University of California, Berkeley}}
\author{Sergey Levine}
\affiliation{\institution{University of California, Berkeley}}
\author{Michiel van de Panne}
\affiliation{\institution{University of British Columbia}}


\setcopyright{acmlicensed}
\acmJournal{TOG}
\acmYear{2018}\acmVolume{37}\acmNumber{4}\acmArticle{143}\acmMonth{8} \acmDOI{10.1145/3197517.3201311}

\begin{document}

\begin{teaserfigure}
 \centering
 \vspace{-0.4cm}
\includegraphics[width=\textwidth]{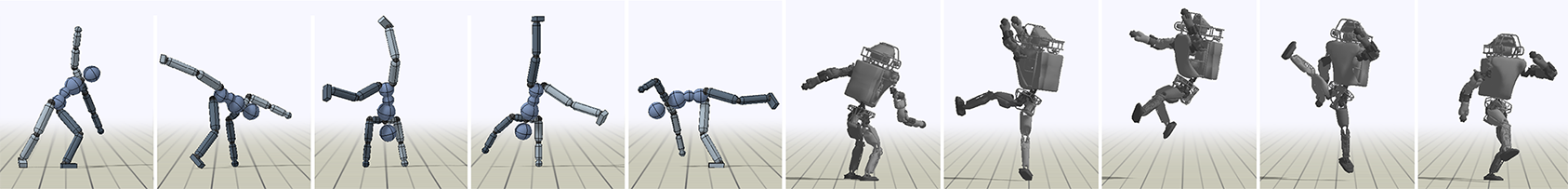}
\vspace{-0.7cm}
   \caption{Highly dynamic skills learned by imitating reference motion capture clips using our method, executed by physically simulated characters. \textbf{Left:} Humanoid character performing a cartwheel. \textbf{Right:} Simulated Atlas robot performing a spinkick.}
   \vspace{-0.05cm}
\end{teaserfigure}

\begin{abstract}
A longstanding goal in character animation is to combine data-driven specification of behavior with a system that can execute a similar behavior in a physical simulation, thus enabling realistic responses to perturbations and environmental variation.
We show that well-known reinforcement learning (RL) methods can be adapted to learn robust control policies 
capable of imitating a broad range of example motion clips, while also learning complex recoveries, 
adapting to changes in morphology, and accomplishing user-specified goals.
Our method handles keyframed motions, highly-dynamic actions such as motion-captured flips and spins, and retargeted motions.
By combining a motion-imitation objective with a task objective, we can train characters that react 
intelligently in interactive settings, e.g., by walking in a desired direction or throwing a ball at a user-specified target.
This approach thus combines the convenience and motion quality of using motion clips to define the desired style and appearance, with the flexibility and generality afforded by RL methods and physics-based animation.
We further explore a number of methods for integrating multiple clips into the learning process to develop 
multi-skilled agents capable of performing a rich repertoire of diverse skills.
We demonstrate results using multiple characters (human, Atlas robot, bipedal dinosaur, dragon) and a large variety of skills, including locomotion, acrobatics, and martial arts.
\vspace{-0.1cm}
\end{abstract}


\maketitle





\section{Introduction}

Physics-based simulation of passive phenomena, such as cloth and fluids, has become nearly ubiquitous in industry. However, the adoption of physically simulated characters has been more modest. Modeling the motion of humans and animals remains a challenging problem, and currently, few methods exist that can simulate the diversity of behaviors exhibited in the real world. Among the enduring challenges in this domain are generalization and directability. Methods that rely on manually designed controllers have produced compelling results, but their ability to generalize to new skills and new situations is limited by the availability of human insight. Though humans are adept at performing a wide range of skills themselves, it can be difficult to articulate the internal strategies that underly such proficiency, and more challenging still to encode them into a controller. Directability is another obstacle that has impeded the adoption of simulated characters. Authoring motions for simulated characters remains notoriously difficult, and current interfaces still cannot provide users with an effective means of eliciting the desired behaviours from simulated characters.

Reinforcement learning (RL) provides a promising approach for motion synthesis, whereby an agent learns to perform various skills through trial-and-error, thus reducing the need for human insight. While deep reinforcement learning
has been demonstrated to produce a range of complex behaviors in prior work~\cite{SchulmanMLJA15,DuanCHSA16,HeessWTLRS16},
the quality of the generated motions has thus far lagged well behind state-of-the-art kinematic methods or manually designed controllers. In particular, controllers trained with deep RL exhibit severe (and sometimes humorous) artifacts, such as extraneous upper body motion, peculiar gaits, and unrealistic posture \cite{HeessTSLMWTEWER17}.\footnote{See, for example, \url{https://youtu.be/hx_bgoTF7bs}}
A natural direction to improve the quality of learned controllers is to incorporate motion capture or hand-authored animation data. In prior work, such systems have typically been designed by layering a physics-based tracking controller on top of a kinematic animation system \cite{DaSilva2008,Lee2010}. This type of approach is challenging because the kinematic animation system must produce reference motions that are feasible to track, and the resulting physics-based controller is limited in its ability to modify the motion to achieve plausible recoveries or accomplish task goals in ways that deviate substantially from the kinematic motion. Furthermore, such methods tend to be quite complex to implement.

An ideal learning-based animation system should allow an artist or motion capture actor to supply a set of reference motions for style, and then generate goal-directed and physically realistic behavior from those reference motions. 
In this work, we take a simple approach to this problem by directly rewarding the learned controller for producing motions that resemble reference animation data, while also achieving additional task objectives.
We also demonstrate three methods for constructing controllers from multiple clips: training with a multi-clip reward based on a {\tt max} operator; 
training a policy to perform multiple diverse skills that can be triggered by the user; and sequencing multiple single-clip policies by using their value functions to estimate the feasibility of transitions.

The central contribution of our paper is a framework for physics-based character animation that combines goal-directed reinforcement learning with data, which may be provided in the form of motion capture clips or keyframed animations. Although our framework consists of individual components that have been known for some time, the particular combination of these components in the context of data-driven and physics-based character animation is novel and, as we demonstrate in our experiments, produces a wide range of skills with motion quality and robustness that substantially exceed prior work.
By incorporating motion capture data into a phase-aware policy, our system can produce physics-based behaviors that are nearly indistinguishable in appearance from the reference motion in the absence of perturbations, avoiding many of the artifacts exhibited by previous deep reinforcement learning algorithms, e.g., \cite{DuanCHSA16}. In the presence of perturbations or modifications, the motions remain natural, and the recovery strategies exhibit a high degree of robustness without the need for human engineering. To the best of our knowledge, we demonstrate some of the most capable physically simulated characters produced by learning-based methods.
In our ablation studies, we identify two specific components of our method, reference state initialization and early termination, that are critical for achieving highly dynamic skills. We also demonstrate several methods for integrating multiple clips into a single policy.

\section{Related Work}

Modeling the skilled movement of articulated figures has a long history in fields ranging from biomechanics to robotics and animation. In recent years, as machine learning algorithms for control have matured, there has also been an increase in interest in these problems from the machine learning community. Here we focus on the most closely related work in animation and RL.

\paragraph{Kinematic Models: } Kinematic methods have been an enduring avenue of work in character animation that can be effective when large amounts of data are available. Given a dataset of motion clips, controllers can be built to select the appropriate clip to play in a given situation, e.g.,~\cite{safonova2007construction,Lee2010MotionFields,Agrawal2016}. Gaussian processes have been used to learn latent representations which can then synthesize motions at runtime \cite{Ye2010,2012-ccclde}. Extending this line of work, deep learning models, such as autoencoders and phase-functioned networks, have also been applied to develop generative models of human motion in a kinematic setting~\cite{Holden2016,Holden2017}. Given high quality data, data-driven kinematic methods will often produce higher quality motions than most simulation-based approaches. However, their ability to synthesize behaviors for novel situations can be limited. As tasks and environments become complex, collecting enough motion data to provide sufficient coverage of the possible behaviors quickly becomes untenable. Incorporating physics as a source of prior knowledge about how motions should change in the presence of perturbations and environmental variation provides one solution to this problem, as discussed below.

\paragraph{Physics-based Models: } Design of controllers for simulated characters remains a challenging problem, and has often relied on human insight to implement task-specific strategies. Locomotion in particular has been the subject of considerable work, with robust controllers being developed for both human and nonhuman characters, e.g., \cite{Yin07,ye2010optimal,2010-TOG-gbwc}. Many such controllers are the products of an underlying simplified model and an optimization process, where a compact set of parameters are tuned in order to achieve the desired behaviors \cite{Wang12optimizinglocomotion,2013-SCA-diverse,HA2014}.
Dynamics-aware optimization methods based on quadratic programming have also been applied to develop locomotion controllers \cite{DaSilva2008,Lee2010,Lee2014}. While model-based methods have been shown to be effective for a variety of skills, they tend to struggle with more dynamics motions that require long-term planning, as well as contact-rich motions. Trajectory optimization has been explored for synthesizing physically-plausible motions for a variety of tasks and characters~\cite{Mordatch2012,Wampler2014}. These methods synthesize motions over an extended time
horizon using an offline optimization process, where the equations of motion are enforced as constraints. 
Recent work has extended such techniques into online model-predictive control methods~\cite{hamalainen2015online,tassa2012synthesis}, although they
remain limited in both motion quality and capacity for long-term planning.
The principal advantage of our method over the above approaches is that of generality.  
We demonstrate that a single model-free framework is capable of a wider range of motion skills 
(from walks to highly dynamic kicks and flips) and an ability to sequence these; 
the ability to combine motion imitation and task-related demands;  compact and fast-to-compute control policies;
and the ability to leverage rich high-dimensional state and environment descriptions.

\paragraph{Reinforcement Learning: } Many of the optimization techniques used to develop controllers for simulated characters are based on reinforcement learning. Value iteration methods have been used to develop kinematic controllers to sequence motion clips in the context of a given task \cite{Lee2010MotionFields,2012-ccclde}. Similar approaches have been explored for simulated characters \cite{Coros09,2015-TOG-terrainRL}.
More recently, the introduction of deep neural network models for RL has given rise to simulated agents that can perform a diverse array of challenging tasks \cite{DuanCHSA16,BrockmanCPSSTZ16,2016-TOG-deepRL,Liu2017,Rajeswaran2017,TehBCQKHHP17}. Policy gradient methods have emerged as the algorithms of choice for many continuous control problems \cite{Sutton1998,SchulmanLMJA15,PPO17}. Although RL algorithms have been capable of synthesizing controllers using minimal task-specific control structures, the resulting behaviors generally appear less natural than their more manually engineered counterparts \cite{SchulmanMLJA15,MerelTTSLWWH17}. Part of the challenge stems from the difficulty in specifying reward functions for natural movement, particularly
in the absence of biomechanical models and objectives that can be used to achieve natural simulated locomotion \cite{Wang12optimizinglocomotion,Lee2014}. Na\"{i}ve objectives for torque-actuated locomotion, such as forward progress or maintaining a desired velocity, often produce gaits that exhibit extraneous motion of the limbs, asymmetric gaits, and other objectionable artifacts. To mitigate these artifacts, additional objectives such as effort or impact penalties have been used to discourage these undesirable behaviors. Crafting such objective functions requires a substantial degree of human insight, and often yields only modest improvements. Alternatively, recent RL methods based on the imitation of motion capture, such as GAIL~\cite{GAIL2016}, address the challenge of designing a reward function by using data to induce an objective. While this has been shown to improve the quality of the generated motions, current results still do not compare favorably to standard methods in computer animation~\cite{MerelTTSLWWH17}. 
The DeepLoco system~\cite{2017-TOG-deepLoco} takes an approach similar to the one we use here, namely adding an imitation term to 
the reward function, although with significant limitations. It uses fixed initial states and is thus not capable of highly dynamic motions;
it is demonstrated only on locomotion tasks defined by foot-placement goals computed by a high-level controller; and it is applied
to a single armless biped model. Lastly, the multi-clip demonstration involves a hand-crafted procedure for 
selecting suitable target clips for turning motions.

\vspace{-0.15cm}
\paragraph{Motion Imitation:} Imitation of reference motions has a long history in computer animation. An early instantiation of this idea was in bipedal locomotion with planar characters \cite{Sharon2005Walking,sok2007simulating}, using controllers tuned through policy search. Model-based methods for tracking reference motions have also been demonstrated for locomotion with 3D humanoid characters \cite{Yin07,muico2009contact,Lee2010}. Reference motions have also been used to shape the reward function for deep RL to produce more natural locomotion gaits \cite{2017-SCA-action,2017-TOG-deepLoco} and for flapping flight \cite{Won2017}. In our work, we demonstrate the capability to perform a significantly broader range
of difficult motions: highly dynamic spins, kicks, and flips with intermittent ground contact, and
we show that reference-state initialization and early termination are critical to their success. We also explore
several options for multi-clip integration and skill sequencing.

The work most reminiscent of ours in terms of capabilities is the Sampling-Based Controller (SAMCON) \cite{2010-TOG-sampControl,2016-TOG-controlGraphs}. 
An impressive array of skills has been reproduced by SAMCON, and to the best of our knowledge, SAMCON has been the only system to demonstrate such a diverse corpus of highly dynamic and acrobatic motions with simulated characters. However, the system is complex, having many components and iterative steps, and requires defining a low dimensional state representation for the synthesized linear feedback structures. The resulting controllers excel at mimicking the original reference motions, but it is not clear how to extend the method for task objectives, particularly if they involve significant sensory input. A more recent variation introduces deep Q-learning to train a high-level policy that selects from a precomputed collection of SAMCON control fragments \cite{Liu2017}. This provides flexibility in the order of execution of the control fragments, and is demonstrated to be capable of challenging non-terminating tasks, such as balancing on a bongo-board and walking on a ball. In this work, we propose an alternative framework using deep RL, that is conceptually much simpler than SAMCON, but is nonetheless able to learn highly dynamic and acrobatic skills, including those having task objectives and multiple clips.

\section{Overview}

Our system receives as input a character model, a corresponding set of kinematic reference motions, and a task defined by a reward function.
It then synthesizes a controller that enables the character to imitate the reference motions, while also satisfying task objectives, such as striking a target or running in a desired direction over irregular terrain. Each reference motion is represented as a sequence of target poses $\{\hat{q}_t\}$. A control policy $\pi(a_t | s_t, g_t)$ maps the state of the character $s_t$, a task-specific goal $g_t$ to an action $a_t$, which is then used to compute torques to be applied to each of the character's joints.
Each action specifies target angles for proportional-derivative (PD) controllers that then produce the final torques applied at the joints.
The reference motions are used to define an imitation reward $r^I(s_t, a_t)$, and the goal defines a task-specific reward $r^G(s_t, a_t, g_t)$.
The final result of our system is a policy that enables a simulated character to imitate the behaviours from the reference motions while also fulfilling the specified task objectives.
The policies are modeled using neural networks and trained using the proximal policy optimization algorithm \cite{PPO17}.

\section{Background}

Our tasks will be structured as standard reinforcement learning problems, where an agent interacts with an environment according to a policy in order to maximize a reward. In the interest of brevity, we will exclude the goal $g$, from the notation, but the following discussion readily generalizes to include this. A policy $\pi(a|s)$ models the conditional distribution over action $a \in A$ given a state $s \in S$. At each control timestep, the agent observes the current state $s_t$ and samples an action $a_t$ from $\pi$. The environment then responds with a new state $s' = s_{t+1}$, sampled from the dynamics $p(s' | s, a)$, and a scalar reward $r_t$ that reflects the desirability of the transition. For a parametric policy $\pi_\theta(a | s)$, the goal of the agent is to learn the optimal parameters $\theta^*$ that maximizes its expected return 
\[J(\theta) = \mathbb{E}_{\tau \sim p_\theta(\tau)} \left[\sum_{t = 0}^T \gamma^t r_t \right],\]
where $p_\theta(\tau) = p(s_0) \prod_{t = 0}^{T - 1} p(s_{t + 1} | s_t, a_t) \pi_\theta(a_t | s_t)$ is the distribution over all possible trajectories $\tau = (s_0, a_0, s_1, ..., a_{T - 1}, s_T)$ induced by the policy $\pi_\theta$, with $p(s_0)$ being the initial state distribution. $\sum_{t = 0}^T \gamma^t r_t$ represents the total return of a trajectory, with a horizon of $T$ steps. $T$ may or may not be infinite, and $\gamma \in [0, 1]$ is a discount factor that can be used to ensure the return is finite. A popular class of algorithms for optimizing a parametric policy is policy gradient methods \citep{sutton2001policy}, where the gradient of the expected return $\triangledown_\theta J(\theta)$ is estimated with trajectories sampled by following the policy. The policy gradient can be estimated according to
\[\triangledown_\theta J(\theta) = \mathbb{E}_{s_t \sim d_\theta(s_t), a_t \sim \pi_\theta(a_t | s_t)} \left[ \triangledown_\theta \mathrm{log}(\pi_\theta(a_t | s_t)) \mathcal{A}_t \right] , \]
where $d_\theta(s_t)$ is the state distribution under the policy $\pi_\theta$. $\mathcal{A}_t$ represents the advantage of taking an action $a_t$ at a given state $s_t$
\[\mathcal{A}_t = R_t - V(s_t) .\]
$R_t = \sum_{l = 0}^{T - t} \gamma^l r_{t + l}$ denotes the return received by a particular trajectory starting from state $s_t$ at time $t$. $V(s_t)$ is a value function that estimates the average return of starting in $s_t$ and following the policy for all subsequent steps
\[V(s_t) = \mathbb{E} \left[ R_t \middle| \pi_\theta, s_t \right] .\]
The policy gradient can therefore be interpreted as increasing the likelihood of actions that lead to higher than expected returns, while decreasing the likelihood of actions that lead to lower than expected returns. A classic policy gradient algorithm for learning a policy using this empirical gradient estimator to perform gradient ascent on $J(\theta)$ is REINFORCE~\citep{Williams1992}.

Our policies will be trained using the proximal policy optimization algorithm \cite{PPO17},
which has demonstrated state-of-the-art results on a number of challenging control problems. The value function will be trained using multi-step returns with TD($\lambda$). The advantages for the policy gradient will be computed using the generalized advantage estimator GAE($\lambda$) \cite{SchulmanMLJA15}.
A more in-depth review of these methods can be found in the supplementary material.

\section{Policy Representation}

Given a reference motion clip, represented by a sequence of target poses $\{\hat{q}_t\}$, the goal of the policy is to reproduce the desired motion in a physically simulated environment, while also satisfying additional task objectives. Since a reference motion only provides kinematic information in the form of target poses, the policy is responsible for determining which actions should be applied at each timestep in order to realize the desired trajectory.

\subsection{States and Actions}

The state $s$ describes the configuration of the character's body, with features consisting of the relative positions of each link with respect to the root (designated to be the pelvis), their rotations expressed in quaternions, and their linear and angular velocities. All features are computed in the character's local coordinate frame, with the root at the origin and the x-axis along the root link's facing direction. Since the target poses from the reference motions vary with time, a phase variable $\phi \in [0, 1]$ is also included among the state features. $\phi = 0$ denotes the start of a motion, and $\phi = 1$ denotes the end. For cyclic motions, $\phi$ is reset to 0 after the end of each cycle.
Policies trained to achieve additional task objectives, such as walking in a particular direction or hitting a target, are also provided with a goal $g$, which can be treated in a similarly fashion as the state. Specific goals used in the experiments are discussed in section~\ref{sec:tasks}.
The action $a$ from the policy specifies target orientations for PD controllers at each joint. The policy is queried at $30 Hz$, and target orientations for spherical joints are represented in axis-angle form, while targets for revolute joints are represented by scalar rotation angles. Unlike the standard benchmarks, which often operate directly on torques, our use of PD controllers abstracts away low-level control details such as local damping and local feedback. Compared to torques, PD controllers have been shown to improve performance and learning speed for certain motion control tasks \cite{2017-SCA-action}.

\subsection{Network}
Each policy $\pi$ is represented by a neural network that maps a given state $s$ and goal $g$ to a distribution over action $\pi(a | s, g)$. The action distribution is modeled as a Gaussian, with a state dependent mean $\mu(s)$ specified by the network, and a fixed diagonal covariance matrix $\Sigma$ that is treated as a hyperparameter of the algorithm:
\[\pi(a | s) = \mathcal{N}(\mu(s), \Sigma) .\]
The inputs are processed by two fully-connected layers with 1024, and 512 units each, followed by a linear output layer. ReLU activations are used for all hidden units. The value function is modeled by a similar network, with exception of the output layer, which consists of a single linear unit.

\begin{figure}[b]
	\centering
    \includegraphics[width=1\columnwidth]{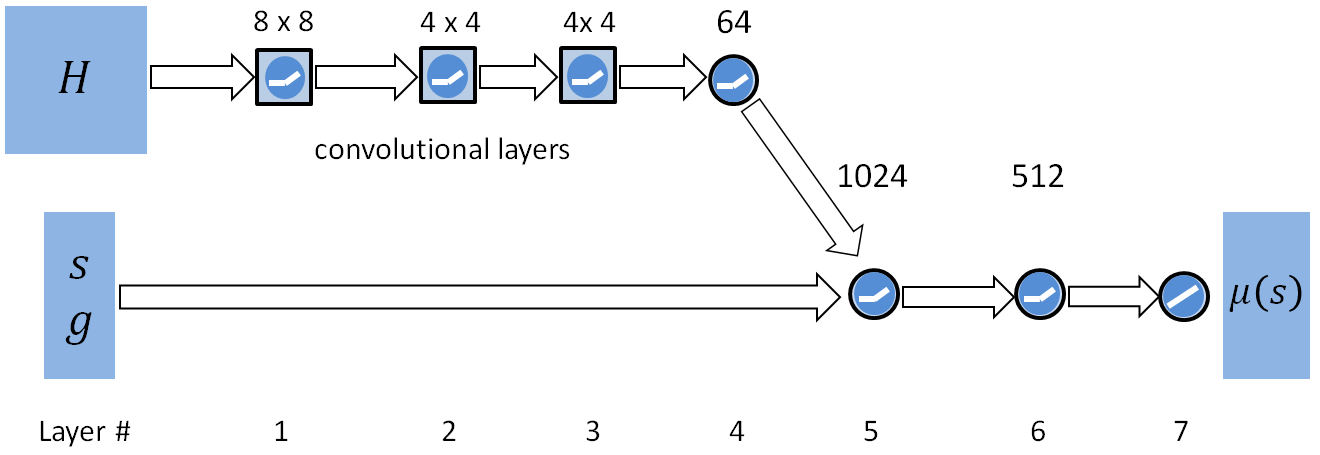}
\caption{Schematic illustration of the visuomotor policy network. The heightmap $H$ is processed by 3 convolutional layers with 16 8x8 filters, 32 4x4 filters, and 32 4x4 filters. The feature maps are then processed by 64 fully-connected units. The resulting features are concatenated with the input state $s$ and goal $g$ and processed by by two fully-connected layer with 1024 and 512 units. The output $\mu(s)$ is produced by a layer of linear units. ReLU activations are used for all hidden layers. For tasks that do not require a heightmap, the networks consist only of layers 5-7.}
\vspace{-0.5cm}
\label{fig:net}
\end{figure}

For vision-based tasks, discussed in section~\ref{sec:tasks},
the inputs are augmented with a heightmap $H$ of the surrounding terrain, sampled on a uniform grid around the character. The policy and value networks are augmented accordingly with convolutional layers to process the heightmap. A schematic illustration of this visuomotor policy network is shown in Figure \ref{fig:net}. The heightmap is first processed by a series of convolutional layers, followed by a fully-connected layer. The resulting features are then concatenated with the input state $s$ and goal $g$, and processed by a similar fully-connected network as the one used for tasks that do not require vision.

\subsection{Reward}
The reward $r_t$ at each step $t$ consists of two terms that encourage the character to match the reference motion while also satisfying additional task objectives:
\[r_t = \omega^I r^I_t + \omega^G r^G_t.\]
Here, $r^I_t$ and $r^G_t$ represent the imitation and task objectives, with $\omega^I$ and $\omega^G$ being their respective weights. The task objective $r^G_t$ incentivizes the character to fulfill task-specific objectives, the details of which will be discussed in the following section. The imitation objective $r^I_t$ encourages the character to follow a given reference motion $\{\hat{q}_t\}$. It is further decomposed into terms that reward the character for matching certain characteristics of the reference motion, such as joint orientations and velocities, as follows:
\\
\[r^I_t = w^p r_t^p + w^v r_t^v + w^e r_t^e + w^c r_t^c\]
\[w^p = 0.65, w^v = 0.1, w^e = 0.15, w^c = 0.1\]
The pose reward $r_t^p$ encourages the character to match the joint orientations of the reference motion at each step, and is computed as the difference between the joint orientation quaternions of the simulated character and those of the reference motion. In the equation below, $q_t^j$ and $\hat{q}_t^j$ represent the orientations of the $j$th joint from the simulated character and reference motion respectively, $q_1 \ominus q_2$ denotes the quaternion difference, and $||q||$ computes the scalar rotation of a quaternion about its axis in radians:
\[r_t^p = \mathrm{exp} \left[-2 \left(\sum_j ||\hat{q}_t^j \ominus q_t^j||^2\right) \right]. \]
The velocity reward $r_t^v$ is computed from the difference of local joint velocities, with $\dot{q}_t^j$ being the angular velocity of the $j$th joint. The target velocity $\hat{\dot{q}}_t^j$ is computed from the data via finite difference. 
\[r_t^v = \mathrm{exp} \left[-0.1 \left(\sum_j ||\hat{\dot{q}}_t^j - \dot{q}_t^j||^2\right) \right]. \]
The end-effector reward $r_t^e$ encourages the character's hands and feet to match the positions from the reference motion. Here, $p_t^e$ denotes the 3D world position in meters of end-effector $e \in [$left foot, right foot, left hand, right hand$]$:
\[r_t^e = \mathrm{exp} \left[-40 \left(\sum_e ||\hat{p}_t^e - p_t^e||^2\right) \right]. \]
Finally, $r_t^c$ penalizes deviations in the character's center-of-mass $p_t^c$ from that of the reference motion $\hat{p}_t^c$:
\[r_t^c = \mathrm{exp} \left[-10 \left(||\hat{p}_t^{c} - p_t^{c}||^2\right) \right]. \]

\section{Training}

Our policies are trained with PPO using the clipped surrogate objective \cite{PPO17}.
We maintain two networks, one for the policy $\pi_\theta(a|s, g)$ and another for the value function $V_\psi(s, g)$, with parameters $\theta$ and $\psi$ respectively. Training proceeds episodically, where at the start of each episode, an initial state $s_0$ is sampled uniformly from the reference motion (section \ref{sec:initState}), and rollouts are generated by sampling actions from the policy at every step. Each episode is simulated to a fixed time horizon or until a termination condition has been triggered (section \ref{sec:earlyTerm}). Once a batch of data has been collected, minibatches are sampled from the dataset and used to update the policy and value function. The value function is updated using target values computed with TD($\lambda$) \cite{Sutton1998}. The policy is updated using gradients computed from the surrogate objective, with advantages $\mathcal{A}_t$ computed using GAE($\lambda$) \cite{SchulmanMLJA15}. Please refer to the supplementary material for a more detailed summary of the learning algorithm.

One of the persistent challenges in RL is the problem of exploration. Since most formulations assume an unknown MDP, the agent is required to use its interactions with the environment to infer the structure of the MDP and discover high value states that it should endeavor to reach. A number of algorithmic improvements have been proposed to improve exploration, such as using metrics for novelty or information gain \cite{BellemareSOSSM16,HouthooftCDSTA16,EX2NIPS2017}.
However, less attention has been placed on the structure of the episodes during training and their potential as a mechanism to guide exploration. In the following sections, we consider two design decisions, the initial state distribution and the termination condition, which have often been treated as fixed properties of a given RL problem. We will show that appropriate choices are crucial for allowing our method to learn challenging skills such as highly-dynamic kicks, spins, and flips.  With common default choices, such as a fixed initial state and fixed-length episodes, we find that imitation of these difficult motions is often unsuccessful.

\vspace{-0.1cm}
\subsection{Initial State Distribution} \label{sec:initState}
The initial state distribution $p(s_0)$ determines the states in which an agent begins each episode. A common choice for $p(s_0)$ is to always place the agent in a fixed state. However, consider the task of imitating a desired motion. A simple strategy is to initialize the character to the starting state of the motion, and allow it to proceed towards the end of the motion over the course of an episode. With this design, the policy must learn the motion in a sequential manner, by first learning the early phases of the motion, and then incrementally progressing towards the later phases. Before mastering the earlier phases, little progress can be made on the later phases. This can be problematic for motions such as backflips, where learning the landing is a prerequisite for the character to receive a high return from the jump itself. If the policy cannot land successfully, jumping will actually result in \emph{worse} returns.
Another disadvantage of a fixed initial state is the resulting exploration challenge. The policy only receives reward retrospectively, once it has visited a state. Therefore, until a high-reward state has been visited, the policy has no way of learning that this state is favorable. Both disadvantages can be mitigated by modifying the initial state distribution.

\begin{figure*}[t]
	\centering
     \subfigure[Humanoid]{   \includegraphics[width=0.2\columnwidth]{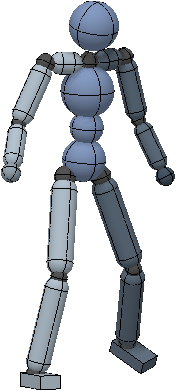}}
     \hspace{2em}
     \subfigure[Atlas]{   \includegraphics[width=0.3\columnwidth]{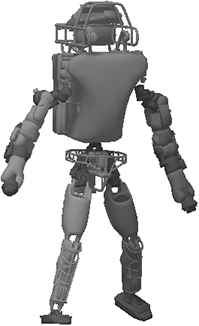}}
     \hspace{1em}
     \subfigure[T-Rex]{   \includegraphics[width=0.62\columnwidth]{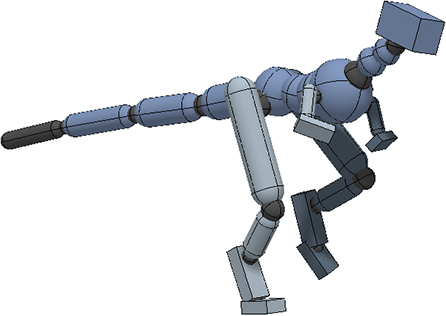}}
	\subfigure[Dragon]{   \includegraphics[width=0.77\columnwidth]{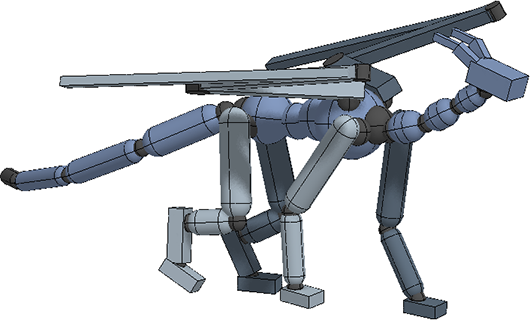}}
    \vspace{-1em}
\caption{3D simulated characters. Our framework is able to train policies for a wide range of character morphologies.}
\label{fig:chars}
\vspace{-0.25cm}
\end{figure*}

For many RL tasks, a fixed initial state can be more convenient, since it can be challenging to initialize the agent in other states (e.g., physical robots) or obtain a richer initial state distribution. For motion imitation tasks, however, the reference motion provides a rich and informative state distribution, that can be leveraged to guide the agent during training. At the start of each episode, a state can be sampled from the reference motion, and used to initialize the state of the agent. We will refer to this strategy as {\em reference state initialization} (RSI). Similar strategies have been previously used for planar bipedal walking \cite{Sharon2005Walking} and manipulation \cite{Nair2017,Rajeswaran2017}.
By sampling initial states from the reference motion, the agent encounters desirable states along the motion, even before the policy has acquired the proficiency needed to reach those states. For example, consider the challenge of learning to perform a backflip. With a fixed initial state, in order for the character to discover that performing a full rotation mid-air will result in high returns, it must first learn to perform a carefully coordinated jump. However, for the character to be motivated to perform such a jump, it must be aware that the jump will lead to states that yield higher rewards. Since the motion is highly sensitive to the initial conditions at takeoff, many strategies will result in failure.
Thus the agent is unlikely to encounter states from a successful flip, and never discover such high reward states. With RSI, the agent immediately encounters such promising states during the early stages of training. Instead of accessing information from the reference motion only through the reward function, RSI can be interpreted as an additional channel through which the agent can access information from the reference motion in the form of a more informative initial state distribution.

\subsection{Early Termination} \label{sec:earlyTerm}

For cyclic skills, the task can be modeled as an infinite horizon MDP. But during training, each episode is simulated for a finite horizon. An episode terminates either after a fixed period of time, or when certain termination conditions have been triggered. For locomotion, a common condition for early termination (ET) is the detection of a fall, characterized by the character's torso making contact with the ground \cite{2016-TOG-deepRL} or certain links falling below a height threshold \cite{HeessWTLRS16}. While these strategies are prevalent, they are often mentioned in passing  and their impact on performance has not been well evaluated. In this work, we will use a similar termination condition as \cite{2016-TOG-deepRL}, where an episode is terminated whenever certain links, such as the torso or head, makes contact with the ground. Once early termination has been triggered, the character is left with zero reward for the remainder of the episode. This instantiation of early termination provides another means of shaping the reward function to discourage undesirable behaviors. Another advantages of early termination is that it can function as a curating mechanism that biases the data distribution in favor of samples that may be more relevant for a task. In the case of skills such as walking and flips, once the character has fallen, it can be challenging for it to recover and return to its nominal trajectory. Without early termination, data collected during the early stages of training will be dominated by samples of the character struggling on the ground in vain, and much of the capacity of the network will be devoted to modeling such futile states. This phenomena is analogous to the class imbalance problem encountered by other methodologies such as supervised learning. By terminating the episode whenever such failure states are encountered, this imbalance can be mitigated.

\section{Multi-Skill Integration} 

The discussion thus far has been focused on imitating individual motion clips. But the ability to compose and sequence multiple clips is vital for performing more complex tasks. In this section, we propose several methods for doing this, which are each suited for different applications. First, we need not be restricted to a single reference clip to define a desired motion style, and can instead choose to use a richer and more flexible {\em multi-clip} reward. Second, we can further provide the user with control over which behavior to trigger, by training a {\em skill selector} policy that takes in a user-specified one-hot clip-selection input. Third, we can avoid training new policies for every clip combination by instead constructing a {\em composite policy} out of existing single-clip policies. In this setup, multiple policies are learned independently and, at runtime, their value functions are used to determine which policy should be activated.

\begin{table}[t]
{ \centering  
\caption{Properties of the characters.}
\label{tab:chars}
\vspace{-0.25cm}
\begin{tabular}{|l|c|c|c|c|}
\hline
{\bf Property} & {\bf Humanoid} & {\bf Atlas} & {\bf T-Rex} & {\bf Dragon} \\ \hline
Links &  13 & 12 & 20 & 32 \\ \hline
Total Mass (kg) & 45 & 169.8 & 54.5 & 72.5 \\ \hline
Height (m) &  1.62 & 1.82 & 1.66 & 1.83 \\ \hline
Degrees of Freedom & 34 & 31 & 55 & 79 \\ \hline
State Features & 197 & 184 & 262 & 418 \\ \hline
Action Parameters & 36 & 32 & 64 & 94 \\ \hline
\end{tabular} \\
}
\vspace{-0.5cm}
\end{table}

\paragraph{Multi-Clip Reward:} To utilize multiple reference motion clips during training, we define a composite imitation objective calculated simply as the max over the previously introduced imitation objective applied to each of the $k$ motion clips:
\[r_t^I = \mathop{\mathrm{max}}_{j = 1,...,k} \ r^{j}_t, \]
where $r^{j}_t$ is the imitation objective with respect to the $j$th clip. We will show that this simple composite objective is sufficient to integrate multiple clips into the learning process. Unlike \cite{2017-TOG-deepLoco}, which required a manually crafted kinematic planner to select a clip for each walking step, our objective provides the policy with the flexibility to select the most appropriately clip for a given situation, and the ability to switch between clips whenever appropriate, without the need to design a kinematic planner.

\paragraph{Skill Selector:} Besides simply providing the policy with multiple clips to use as needed to accomplish a goal, we can also provide the user with control over which clip to use at any given time. In this approach, we can train a policy that simultaneously learns to imitate a set of diverse skills and, once trained, is able to execute arbitrary sequences of skills on demand. The policy is provided with a goal $g_t$ represented by a one-hot vector, where each entry $g_{t, i}$ corresponds to the motion that should be executed. The character's goal then is to perform the motion corresponding to the nonzero entry of $g_t$. There is no additional task objective $r^G_t$, and the character is trained only to optimize the imitation objective $r^I_t$, which is computed based on the currently selected motion $r^I_t = r^i_t$, where $g_{t, i} = 1$ and $g_{t, j} = 0$ for $j \not= i$. During training, a random $g_t$ is sampled at the start of each cycle. The policy is therefore required to learn to transition between all skills within the set of clips.

\paragraph{Composite Policy:} The previously described methods both learn a single policy for a collection of clips. But requiring a network to  learn multiple skills jointly can be challenging as the number of skills increases, and can result in the policy failing to learn any of the skills adequately.
An alternative is to adopt a divide-and-conquer strategy, where separate policies are trained to perform different skills, and then integrated together into a composite policy. Since the value function provides an estimate of a policy's expected performance in a particular state, the value functions can be leveraged to determine the most appropriate skill to execute in a given state. Given a set of policies and their value functions $\{V^i(s), \pi^i(a | s)\}_{i = 1}^k$, a composite policy $\Pi(a | s)$ can be constructed using a Boltzmann distribution
\[\Pi(a | s) = \sum_{i=1}^k p^i(s) \pi^i(a|s), \quad p^i(s) = \frac{\mathrm{exp}\left[ V^i(s) / \mathcal{T} \right]}{\sum_{j=1}^k \mathrm{exp}\left[ V^j(s) / \mathcal{T} \right]} , \]
where $\mathcal{T}$ is a temperature parameter. Policies with larger expected values at a given state will therefore be more likely to be selected. By repeatedly sampling from the composite policy, the character is then able to perform sequences of skills from a library of diverse motions without requiring any additional training. The composite policy resembles the mixture of actor-critics experts model (MACE) proposed by \cite{2016-TOG-deepRL}, although it is even simpler as each sub-policy is trained independently for a specific skill.

\section{Characters}

Our characters include a 3D humanoid, an Atlas robot model, a T-Rex, and a dragon.
Illustrations of the characters are available in Figure \ref{fig:chars}, and Table \ref{tab:chars} details the properties of each character. All characters are modeled as articulated rigid bodies, with each link attached to its parent link via a 3 degree-of-freedom spherical joint, except for the knees and elbows, which are attached via 1 degree-of-freedom revolute joints. PD controllers are positioned at each joint, with manually specified gains that are kept constant across all tasks. Both the humanoid and Atlas share similar body structures, but their morphology (e.g., mass distribution) and actuators (e.g., PD gains and torque limits) differ significantly, with the Atlas being almost four times the mass of the humanoid. The T-Rex and dragon provide examples of learning behaviors for characters from keyframed animation when no mocap data is available, and illustrate that our method can be readily applied to non-bipedal characters. The humanoid character has a 197D state space and a 36D action space. Our most complex character, the dragon, has a 418D state space and 94D action space. Compared to standard continuous control benchmarks for RL \cite{OpenAIGym}, which typically have action spaces varying between 3D to 17D, our characters have significantly higher-dimensional action spaces. 

\section{Tasks}\label{sec:tasks}
In addition to imitating a set of motion clips, the policies can 
be trained to perform a variety of tasks while preserving the style prescribed by the reference motions. The task-specific behaviors are encoded into the task objective $r^G_t$. We describe the tasks evaluated in our experiments below.

\begin{figure}[t]
	\centering
    \subfigure{\includegraphics[width=1\columnwidth]{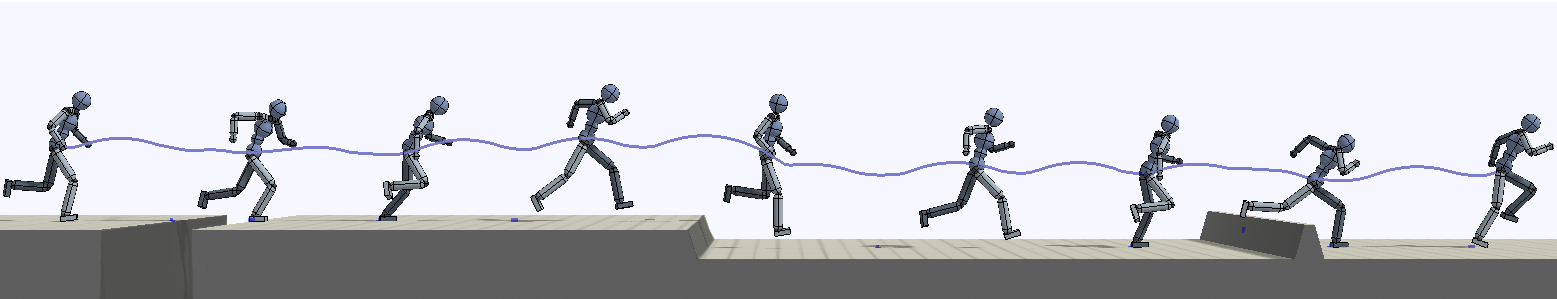}}\\
    \vspace{-1em}
    \subfigure{\includegraphics[width=1\columnwidth]{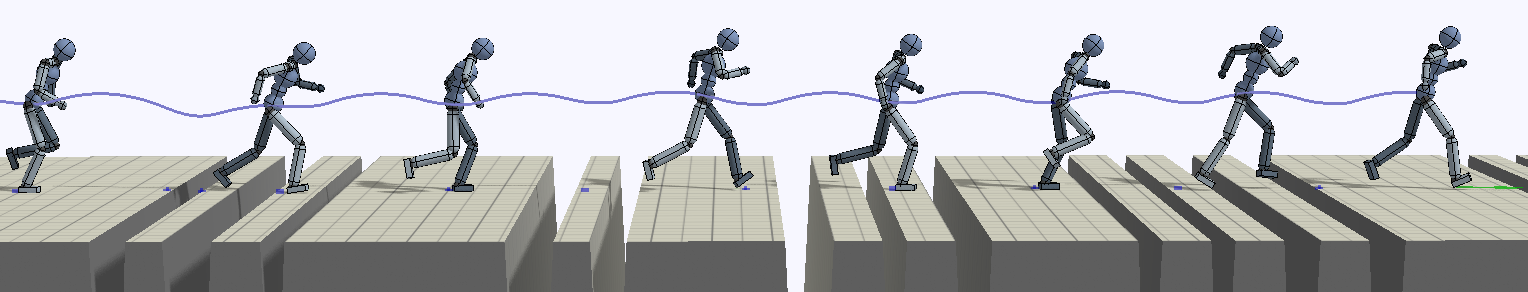}}\\
    \vspace{-1em}
    \subfigure{\includegraphics[width=1\columnwidth]{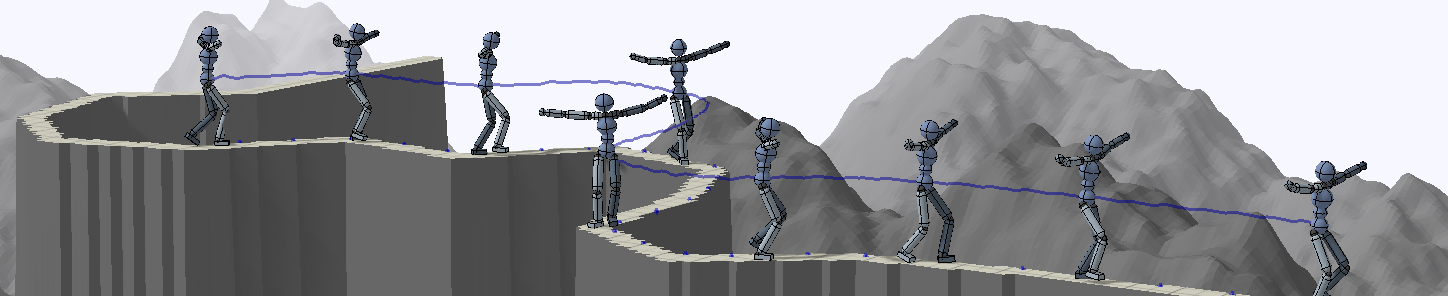}}\\
    \vspace{-1em}
    \subfigure{\includegraphics[width=1\columnwidth]{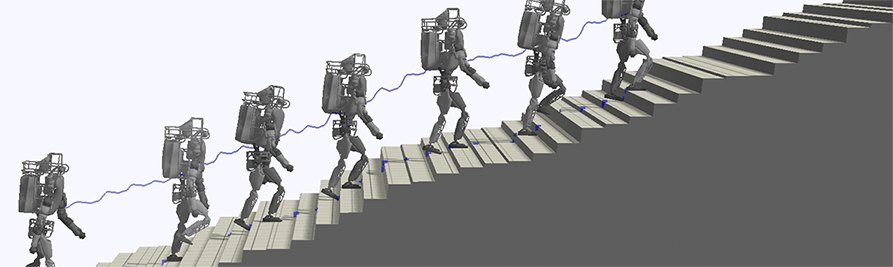}}\\
    \vspace{-1em}
\caption{Characters traversing randomly generated terrains. \textbf{Top-to-bottom:} mixed obstacles, dense gaps, winding balance beam, stairs. The blue line traces the trajectory of the character's center-of-mass.}
\label{fig:terrains}
\vspace{-0.45cm}
\end{figure}

\paragraph{Target Heading:} Steerable controllers can be trained by introducing an objective that encourages the character to travel in a target direction $d^*_t$, represented as a 2D unit vector in the horizontal plane. The reward for this task is given by
\[r^G_t = \mathrm{exp}\left[-2.5 \ \mathrm{max}(0, v^* - v_t^T d^*_t)^2 \right] , \]
where $v^*$ specifies the desired speed along the target direction $d^*_t$, and $v_t$ represents the center-of-mass velocity of the simulated character. The objective therefore penalizes the character for traveling slower than the desired speed along the target direction, but does not penalize it for exceeding the requested speed. The target direction is provided as the input goal $g_t = d^*_t$ to the policy. During training, the target direction is randomly varied throughout an episode. At runtime, $d^*_t$ can be manually specified to steer the character.

\begin{table}[t]
{ \centering  
\caption{Performance statistics of imitating various skills. All skills are performed by the humanoid unless stated otherwise. Policies are trained only to imitate a reference motion without additional task objectives. $T_{cycle}$ is the length of the clip. $N_{samples}$ specifies the number of samples collected to train the final policy. $NR$ represents the normalized return of the final policy averaged over 32 episodes, with 0 being the minimum possible return per episode, and 1 the maximum return. For cyclic skills, each episode has a horizon of 20$s$. For acyclic skills, the horizon is specified by $T_{cycle}$.}
\label{tab:perf}
\vspace{-0.25cm}
\begin{tabular}{|l|c|c|c|}
\hline
{\bf Skill} & {\bf $\boldsymbol T_{cycle}$ (s)} & {\bf $\boldsymbol N_{samples}$} ($\boldsymbol 10^6$) & {\bf $\boldsymbol NR$} \\ \hline
Backflip & 1.75 & 72 & 0.729 \\ \hline
Balance Beam & 0.73 & 96 & 0.783 \\ \hline
Baseball Pitch & 2.47 & 57 & 0.785 \\ \hline
Cartwheel & 2.72 & 51 & 0.804 \\ \hline
Crawl & 2.93 & 68 & 0.932 \\ \hline
Dance A & 1.62 & 67 & 0.863 \\ \hline
Dance B & 2.53 & 79 & 0.822 \\ \hline
Frontflip & 1.65 & 81 & 0.485 \\ \hline
Getup Face-Down & 3.28 & 49 & 0.885 \\ \hline
Getup Face-Up & 4.02 & 66 & 0.838 \\ \hline
Headspin & 1.92 & 112 & 0.640 \\ \hline
Jog & 0.80 & 51 & 0.951 \\ \hline
Jump & 1.77 & 86 & 0.947 \\ \hline
Kick & 1.53 & 50 & 0.854 \\ \hline
Landing & 2.83 & 66 & 0.590 \\ \hline
Punch & 2.13 & 60 & 0.812 \\ \hline
Roll & 2.02 & 81 & 0.735 \\ \hline
Run & 0.80 & 53 & 0.951 \\ \hline
Sideflip & 2.44 & 64 & 0.805 \\ \hline
Spin & 4.42 & 191 & 0.664 \\ \hline
Spinkick & 1.28 & 67 & 0.748 \\ \hline
Vault 1-Handed & 1.53 & 41 & 0.695 \\ \hline
Vault 2-Handed & 1.90 & 87 & 0.757 \\ \hline
Walk & 1.26 & 61 & 0.985 \\ \hline
Atlas: Backflip & 1.75 & 63 & 0.630 \\ \hline
Atlas: Run & 0.80 & 48 & 0.846 \\ \hline
Atlas: Spinkick & 1.28 & 66 & 0.477 \\ \hline
Atlas: Walk & 1.26 & 44 & 0.988 \\ \hline
T-Rex: Walk & 2.00 & 140 & 0.979 \\ \hline
Dragon: Walk & 1.50 & 139 & 0.990 \\ \hline
\end{tabular} \\
}
\vspace{-0.4cm}
\end{table}

\paragraph{Strike:} In this task, the character's goal is to strike a randomly placed spherical target using specific links, such as the feet. The reward is given by
\[r^G_t = 
\begin{cases}
1, & \text{target has been hit} \\
\mathrm{exp}\left[-4 ||p^{tar}_t - p^e_t||^2 \right], & \text{otherwise} \\
\end{cases}
\]
$p^{tar}_t$ specifies the location of the target, and $p^e_t$ represents the position of the link used to hit the target. The target is marked as being hit if the center of the link is within 0.2m of the target location. The goal $g_t = (p^{tar}_t, h)$ consists of the target location $p^{tar}_t$ and a binary variable $h$ that indicates if the target has been hit in a previous timestep. As we are using feedforward networks for all policies, $h$ acts as a memory for the state of the target. The target is randomly placed within a distance of [0.6, 0.8]$m$ from the character, the height is sampled randomly between between [0.8, 1.25]$m$, and the initial direction from the character to the target varies by 2$rad$. The target location and $h$ are reset at the start of each cycle. The memory state $h$ can be removed by training a recurrent policy, but our simple solution avoids the complexities of training recurrent networks while still attaining good performance.

\begin{figure}[t]
	\centering
    \subfigure{\includegraphics[width=1\columnwidth]{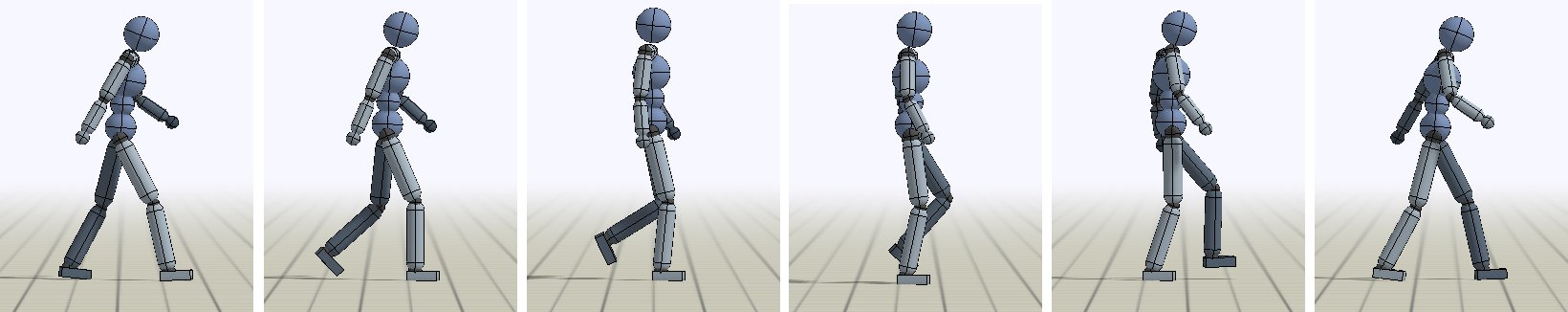}} \\
    \vspace{-1em}
    \subfigure{\includegraphics[width=1\columnwidth]{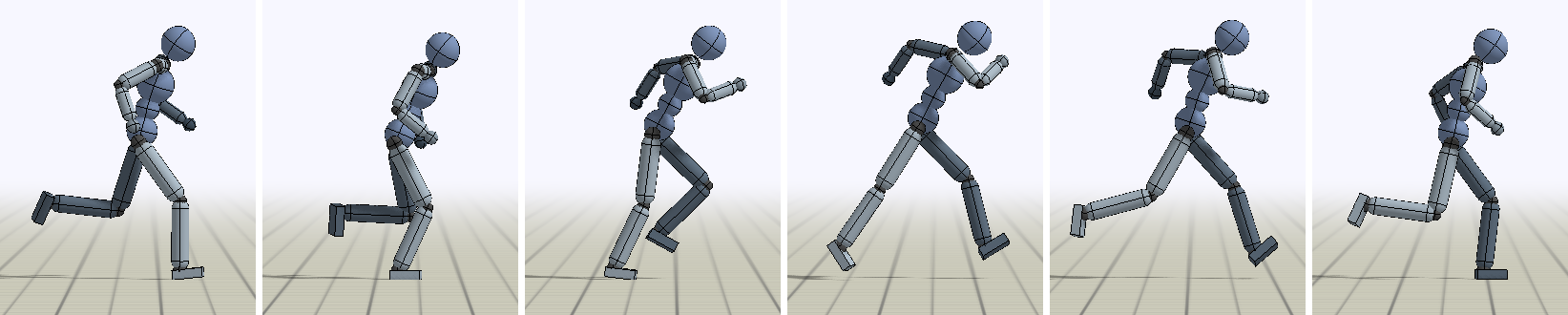}} \\
    \vspace{-1em}
    \subfigure{\includegraphics[width=1\columnwidth]{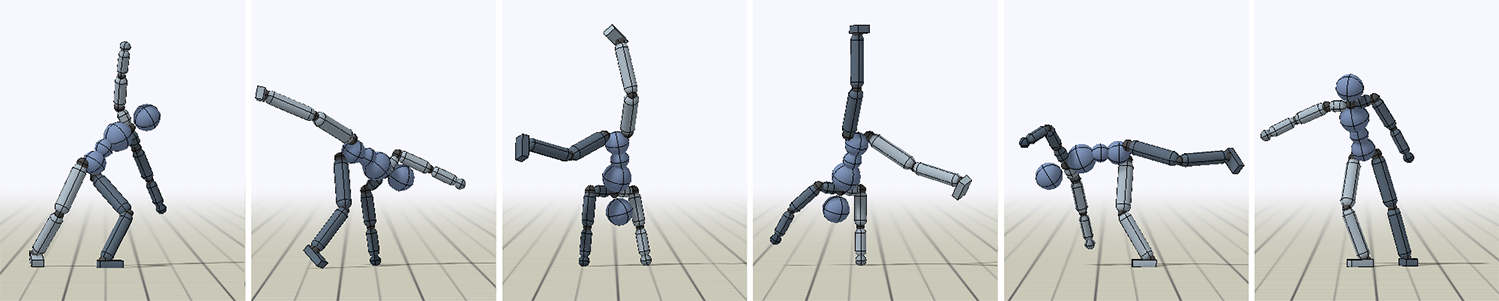}} \\
    \vspace{-1em}
    \subfigure{\includegraphics[width=1\columnwidth]{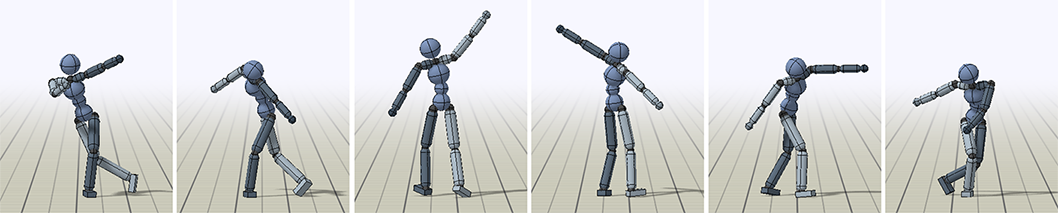}} \\
    \vspace{-1em}
    \subfigure{\includegraphics[width=1\columnwidth]{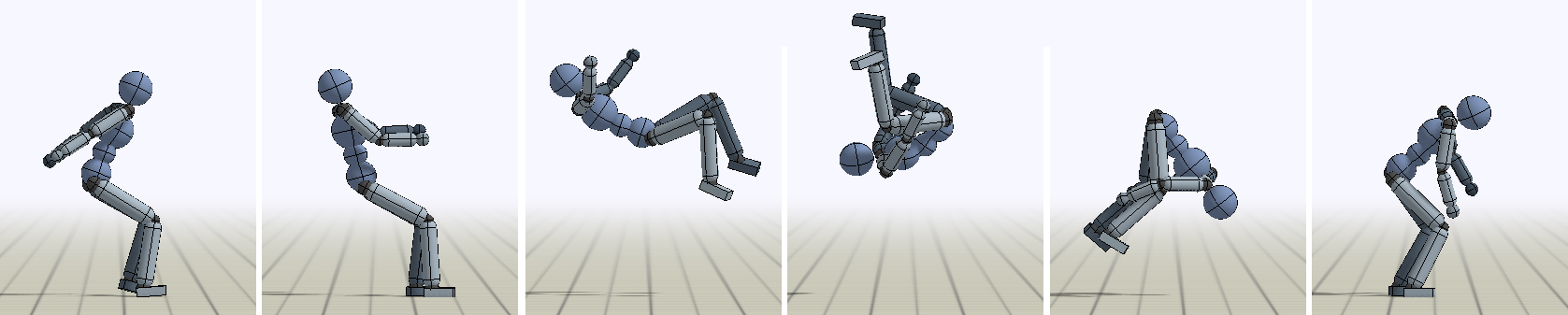}} \\
    \vspace{-1em}
    \subfigure{\includegraphics[width=1\columnwidth]{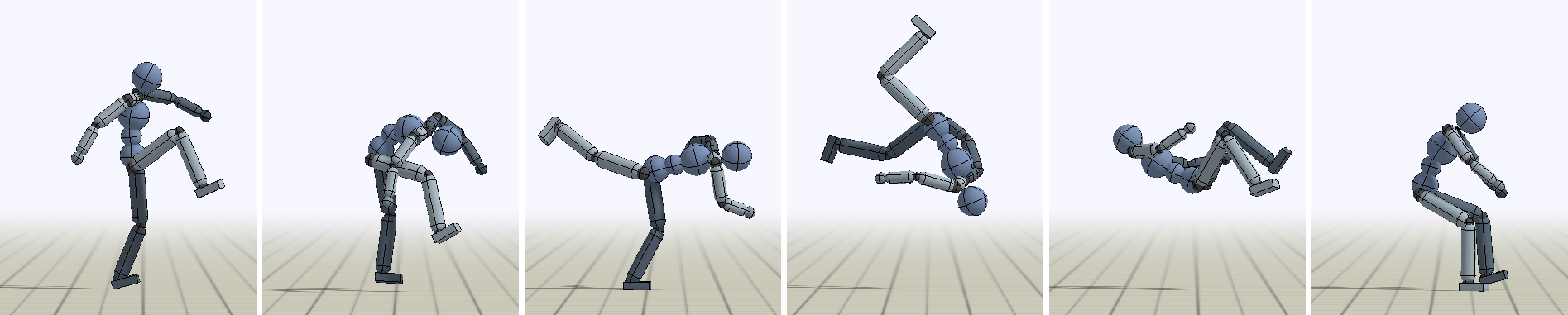}} \\
    \vspace{-1em}
    \subfigure{\includegraphics[width=1\columnwidth]{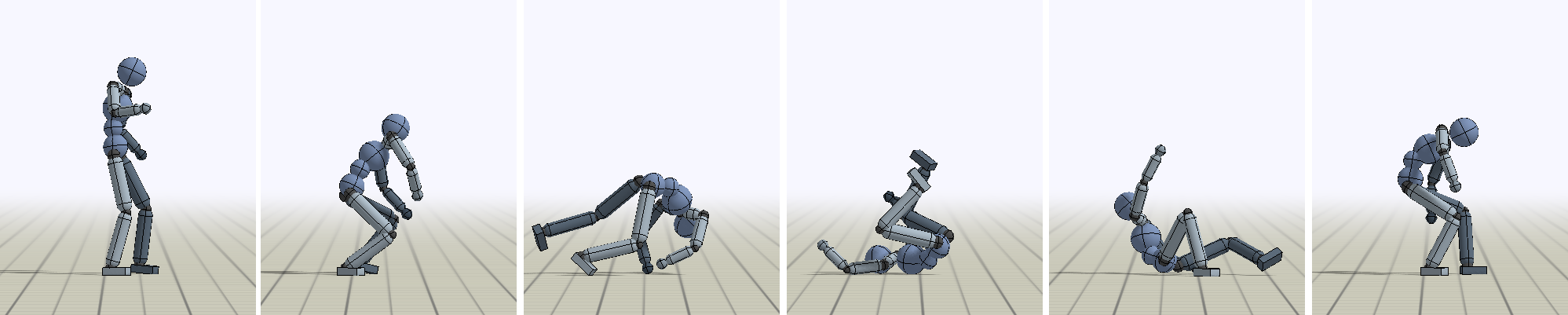}} \\
    \vspace{-1em}
\caption{Snapshots of motions from the trained policies. \textbf{Top-to-bottom:} walk, run, cartwheel, dance A, backflip, frontflip, roll.}
\label{fig:snapshots0}
\vspace{-0.5cm}
\end{figure}

\begin{figure*}[t!]
	\centering
     \subfigure[Atlas: Walk]{   \includegraphics[width=1.03\columnwidth]{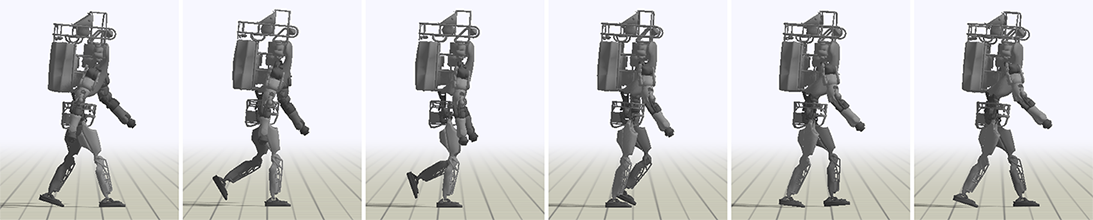}}
     \subfigure[Atlas: Run]{   \includegraphics[width=1.03\columnwidth]{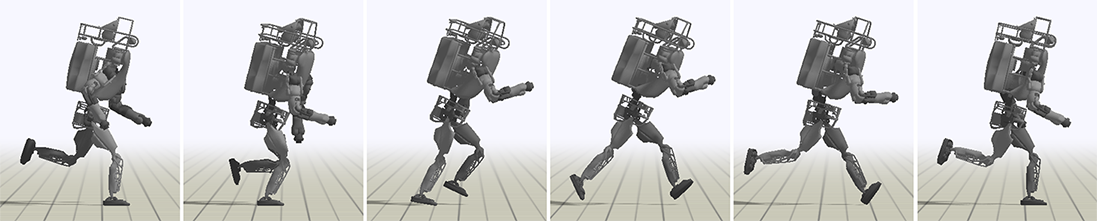}}\\
     \vspace{-1em}
     \subfigure[Atlas: Backflip]{   \includegraphics[width=1.03\columnwidth]{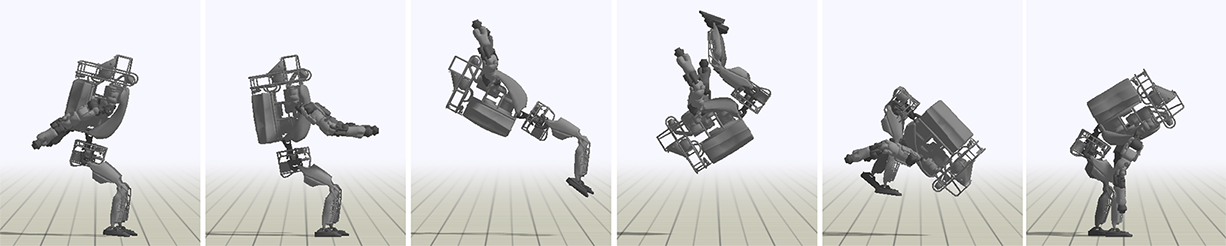}}
     \subfigure[Atlas: Spinkick]{   \includegraphics[width=1.03\columnwidth]{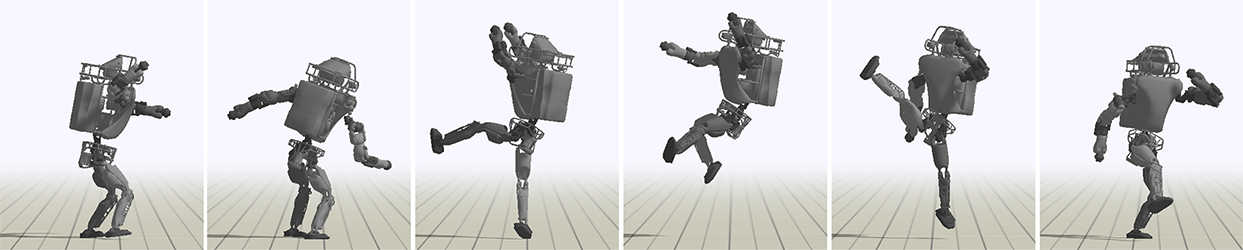}}\\
     \subfigure[T-Rex: Walk]{   \includegraphics[width=1\textwidth]{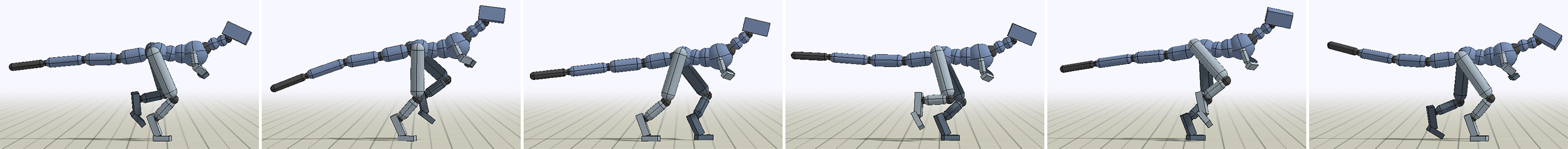}}\\
     \subfigure[Dragon: Walk]{   \includegraphics[width=1\textwidth]{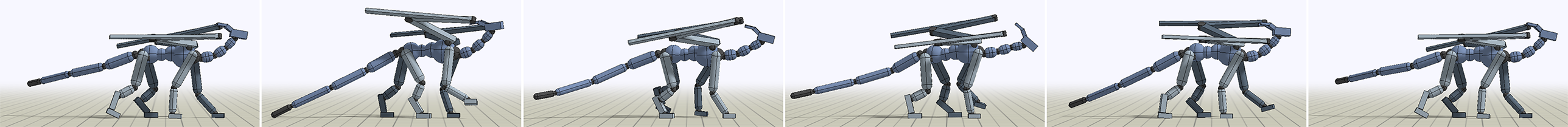}}\\
     \subfigure[Humanoid: Sideflip]{\includegraphics[width=1.04\columnwidth]{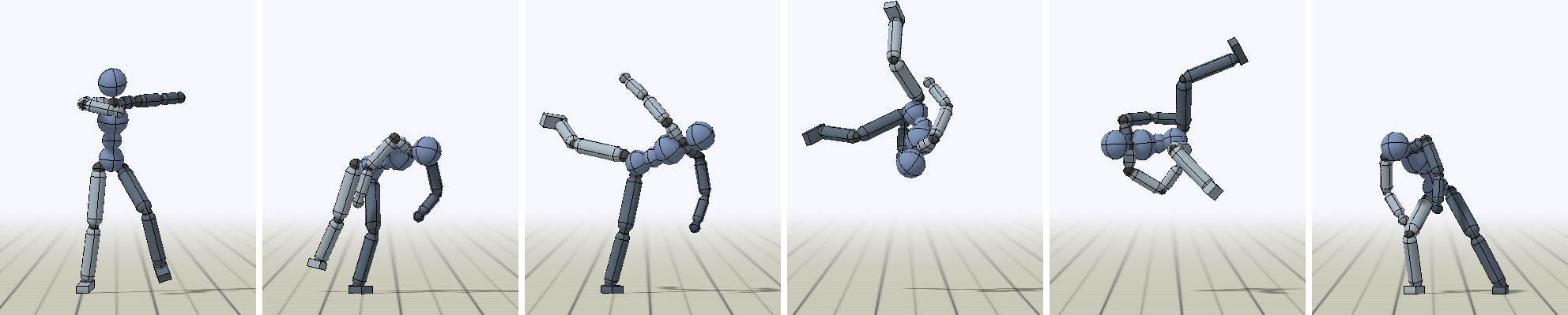}}
    \hspace{0.002\columnwidth}
     \subfigure[Humanoid: Spin]{\includegraphics[width=1.04\columnwidth]{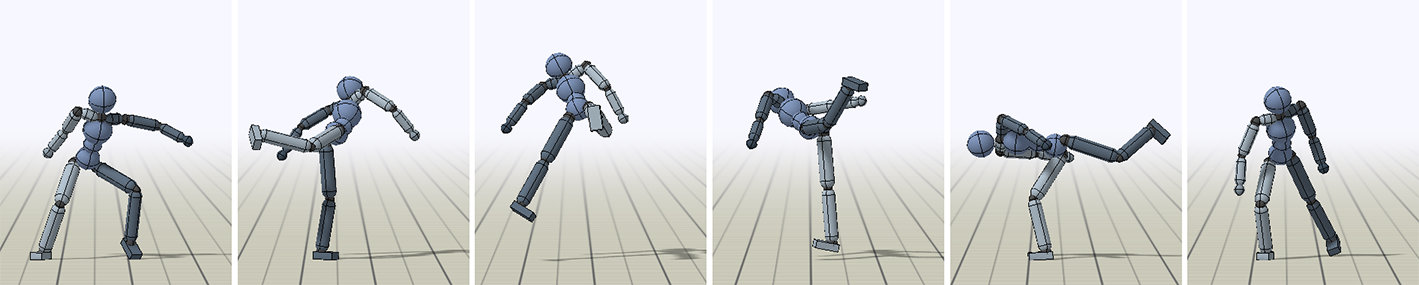}}
     \subfigure[Humanoid: Getup Face-Down]{\includegraphics[width=1.04\columnwidth]{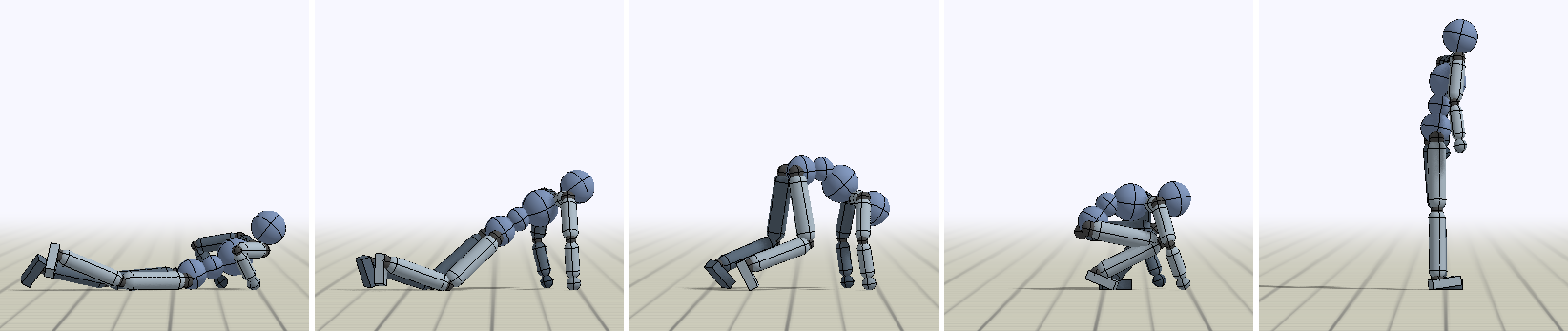}}
    \hspace{0.002\columnwidth}
     \subfigure[Humanoid: Getup Face-Up]{\includegraphics[width=1.04\columnwidth]{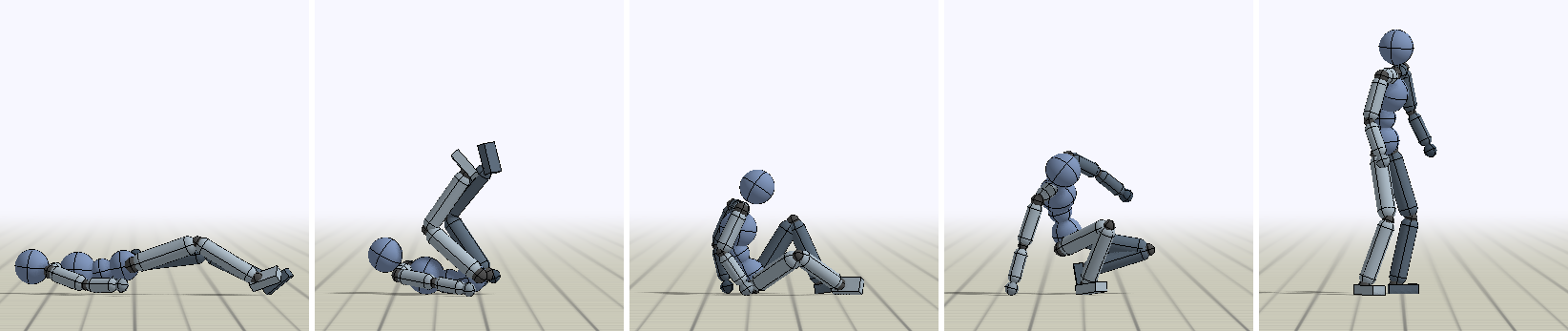}} \\
     \subfigure[Humanoid: Kick]{\includegraphics[width=0.98\columnwidth]{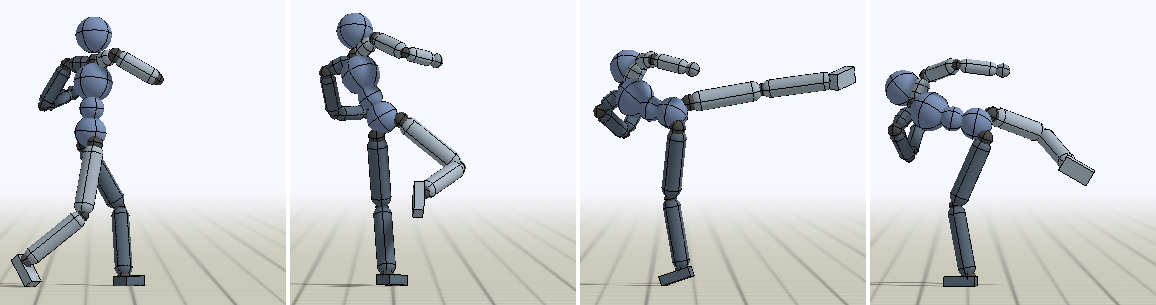}}
    \hspace{0.002\columnwidth}
     \subfigure[Humanoid: Vault 1-Handed]{\includegraphics[width=0.62\columnwidth]{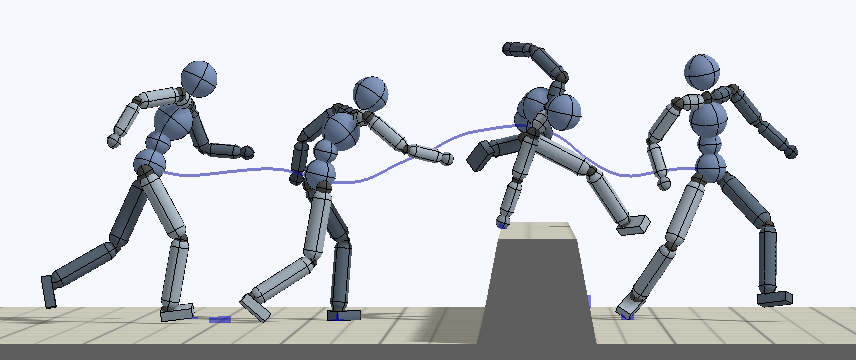}}
    \hspace{0.002\columnwidth}
     \subfigure[Humanoid: Vault 2-Handed]{\includegraphics[width=0.45\columnwidth]{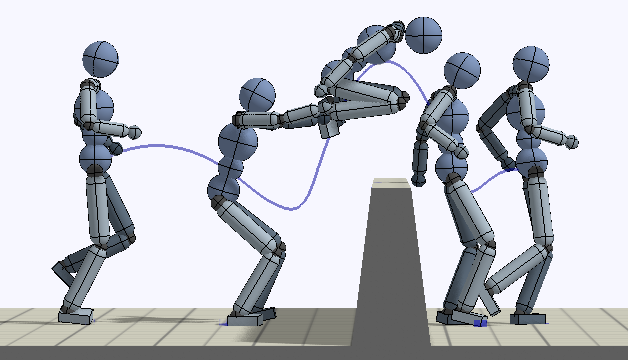}} \\
     \subfigure[Humanoid: Punch]{\includegraphics[width=0.98\columnwidth]{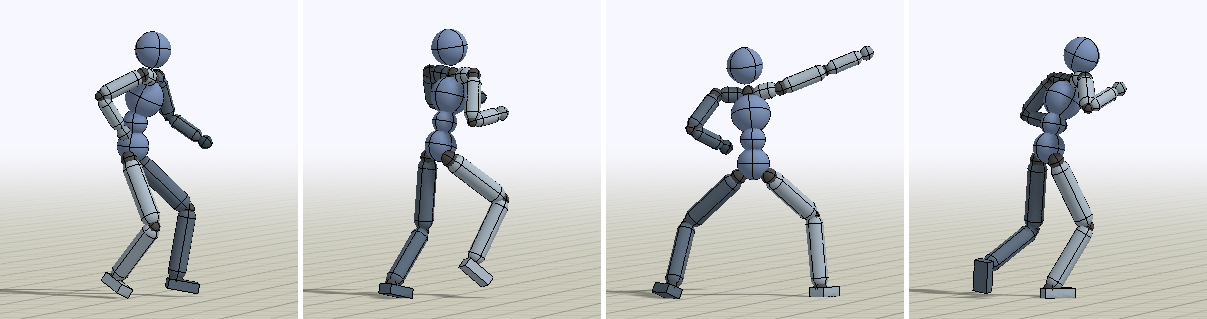}}
    \hspace{0.005\columnwidth}
     \subfigure[Humanoid: Crawl]{\includegraphics[width=1.08\columnwidth]{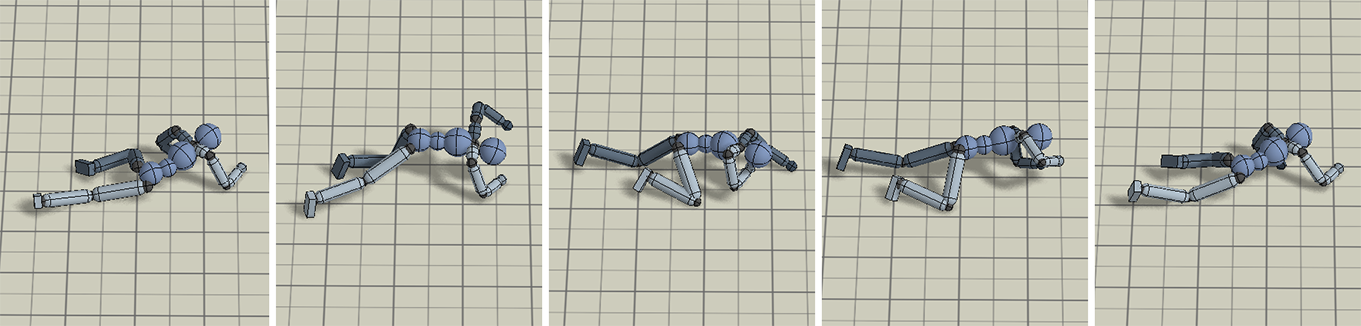}} \\
     \vspace{-1em}
\caption{Simulated characters performing various skills. Our framework is able to train policies for a broad range of characters, skills, and environments.}
\label{fig:snapshots1}
\end{figure*}

\paragraph{Throw:} This task is a variant of the strike task, where instead of hitting a target with one of the character's links, the character is tasked with throwing a ball to the target. At the start of an episode, the ball is attached to the character's hand via a spherical joint. The joint is released at a fixed point in time during the episode. The goal $g_t$ and reward $r^G_t$ is the same as the strike task, but the character state $s_t$ is augmented with the position, rotation, linear and angular velocity of the ball. The distance of the target varies between [2.5, 3.5]$m$, with height between [1, 1.25]$m$, and direction direction between [0.7, 0.9]$rad$.

\paragraph{Terrain Traversal:} In this task, the character is trained to traverse obstacle-filled environments. The goal $g_t$ and task-objective $r^G_t$ are similar to those of the target heading task, except the target heading is fixed along the direction of forward progress.

We consider four environments consisting of mixed obstacles, dense gaps, a winding balance beam, and stairs. Figure~\ref{fig:terrains} illustrates examples of the environments. The mixed obstacles environment is composed of gap, step, and wall obstacles similar to those presented in \cite{2016-TOG-deepRL}. Each gap has a width between [0.2, 1]$m$, the height of each wall varies between [0.25, 0.4]$m$, and each step has a height between [0.35, -0.35]$m$. The obstacles are interleaved with flat stretches of terrain between [5, 8]$m$ in length. The next environment consists of sequences of densely packed gaps, where each sequence consists of 1 to 4 gaps. The gaps are [0.1, 0.3]$m$ in width, with [0.2, 0.4]$m$ of separation between adjacent gaps. Sequences of gaps are separate by [1, 2]$m$ of flat terrain. The winding balance beam environment sports a narrow winding path carved into irregular terrain. The width of the path is approximately 0.4$m$. Finally, we constructed a stairs environment, where the character is to climb up irregular steps with height varying between [0.01, 0.2]$m$ and a depth of 0.28$m$.

To facilitate faster training, we adopt a progressive learning approach, where the standard fully-connected networks (i.e., without the input heightmap and convolutional layers) are first trained to imitate their respective motions on flat terrain. Next, the networks are augmented with an input heightmap and corresponding convolutional layers, then trained in the irregular environments. Since the mixed obstacles and dense gaps environments follow a linear layout, the heightmaps are represented by a 1D heightfield with 100 samples spanning 10$m$. In the winding balance beam environment, a $32 \times 32$ heightmap is used, covering a $3.5\times3.5 m$ area.

\begin{figure}[t]
	\centering
    \subfigure{\includegraphics[width=1\columnwidth]{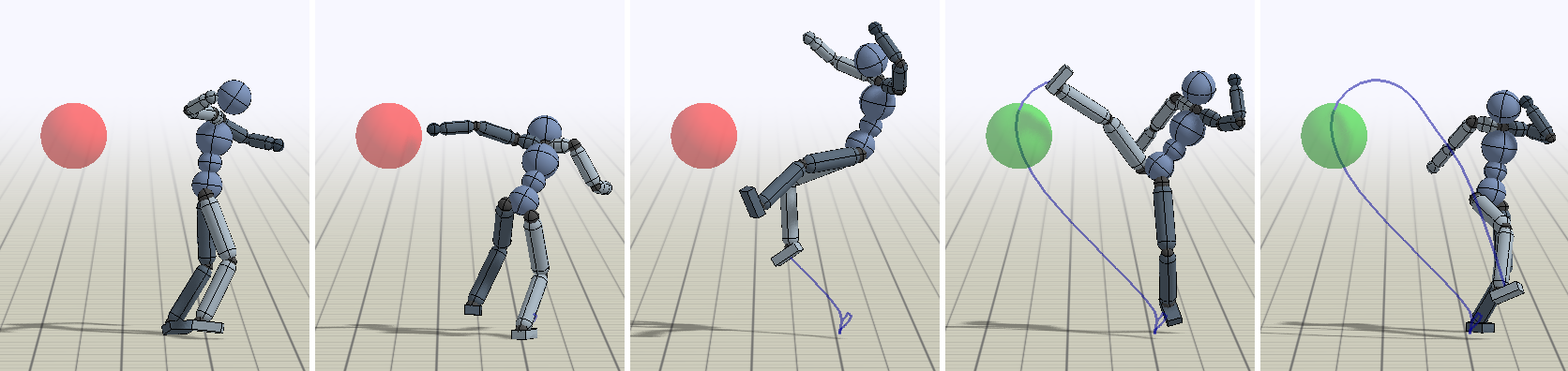}}\\
    \vspace{-0.5em}
    \subfigure{\includegraphics[width=1\columnwidth]{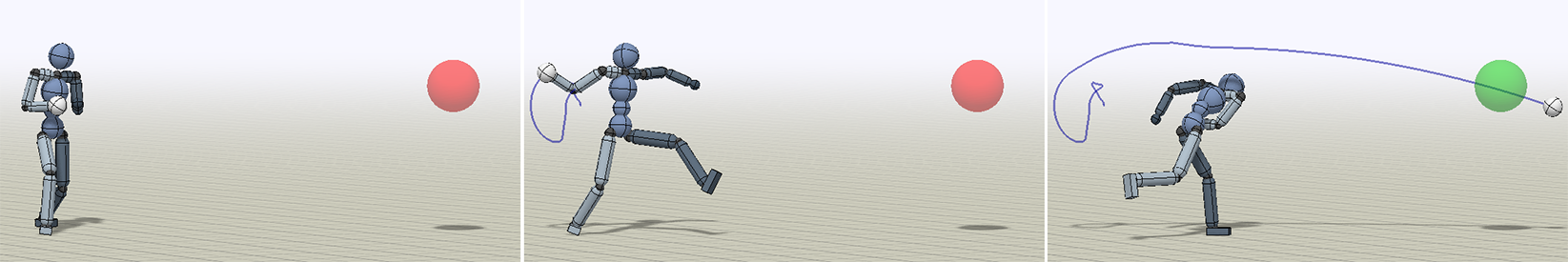}}\\
    \vspace{-1em}
\caption{\textbf{Top:} Spinkick policy trained to strike a target with the character's right foot. \textbf{Bottom:} Baseball pitch policy trained to throw a ball to a target.}
\label{fig:strikeThrow}
\vspace{-0.5cm}
\end{figure}

\section{Results}
The motions from the trained policies are best seen in the supplemental videos.
Snapshots of the skills performed by the simulated characters are available in Figure \ref{fig:snapshots0}, \ref{fig:snapshots1}, and \ref{fig:strikeThrow}. The policies are executed at 30Hz. Physics simulation is performed at 1.2kHz using the Bullet physics engine \citep{Bullet}. All neural networks are built and trained using TensorFlow. The characters' motions are driven by torques computed using stable PD controllers \cite{Tan:2011b}. Results for the humanoid are demonstrated for a large collection of locomotion, acrobatic, and martial arts skills, while the results for the dragon and T-Rex are demonstrated for locomotion. Each skill is learned from approximately 0.5-5s of mocap data collected from \url{http://mocap.cs.cmu.edu} and \url{http://mocap.cs.sfu.ca}. For characters such as the T-Rex and dragon, where mocap data is not available, we demonstrate that our framework is also capable of learning skills from artist-authored keyframes. Before being used for training, the clips are manually processed and retargeted to their respective characters. A comprehensive list of the learned skills and performance statistics is available in Table \ref{tab:perf}. Learning curves for all policies are available in the supplemental material. Each environment is denoted by ``Character: Skill - Task''. The task is left unspecified for policies that are trained solely to imitate a reference motion without additional task objectives.
Performance is measured by the average return normalized by the minimum and maximum possible return per episode. Note that the maximum return may not be achievable. For example, for the throwing task, the maximum return requires moving the ball instantaneously to the target. When evaluating the performance of the policies,
early termination is applied and the state of the character at the start of each episode is initialized via RSI. The weights for the imitation and task objectives are set to $\omega^I = 0.7$ and $\omega^G = 0.3$ for all tasks. More details regarding hyperparameter settings are available in the supplemental material.

\begin{table}[t]
{ \centering  
\caption{Performance statistics of imitating motion clips while also fulfilling additional task objectives.}
\label{tab:taskPerf}
\vspace{-0.25cm}
\begin{tabular}{|l|c|c|}
\hline
{\bf Environment} & {\bf $\boldsymbol N_{samples}$} ($\boldsymbol 10^6$) & {\bf $\boldsymbol NR$} \\ \hline
Humanoid: Walk - Target Heading & 85 & 0.911 \\ \hline
Humanoid: Jog - Target Heading & 108 & 0.876 \\ \hline
Humanoid: Run - Target Heading & 40 & 0.637 \\ \hline
Humanoid: Spinkick - Strike & 85 & 0.601 \\ \hline
Humanoid: Baseball Pitch - Throw & 221 & 0.675 \\ \hline
Humanoid: Run - Mixed Obstacles & 466 & 0.285 \\ \hline
Humanoid: Run - Dense Gaps & 265 & 0.650 \\ \hline
Humanoid: Winding Balance Beam & 124 & 0.439 \\ \hline
Atlas: Walk - Stairs & 174 & 0.808 \\ \hline
\end{tabular} \\
}
\vspace{-0.5cm}
\end{table}

By encouraging the policies to imitate motion capture data from human subjects, our system is able to learn policies for a rich repertoire of skills. For locomotion skills such as walking and running, our policies produce natural gaits that avoid many of the artifacts exhibited by previous deep RL methods \cite{SchulmanMLJA15,MerelTTSLWWH17}. The humanoid is able to learn a variety of acrobatic skills with long flight phases, such as backflips and spinkicks, which are are particularly challenging since the character needs to learn to coordinate its motion in mid-air. The system is also able to reproduce contact-rich motions, such as crawling and rolling, as well as motions that require coordinated interaction with the environment, such as the vaulting skills shown in Figure~\ref{fig:snapshots1}.
The learned policies are robust to significant external perturbation and generate plausible recovery behaviors. The policies trained for the T-Rex and dragon demonstrate that the system can also learn from artist generated keyframes when mocap is not available and scale to much more complex characters than those that have been demonstrated by previous work using deep RL.

\subsection{Tasks}
\begin{figure}[t]
	\centering
    \includegraphics[width=1\columnwidth]{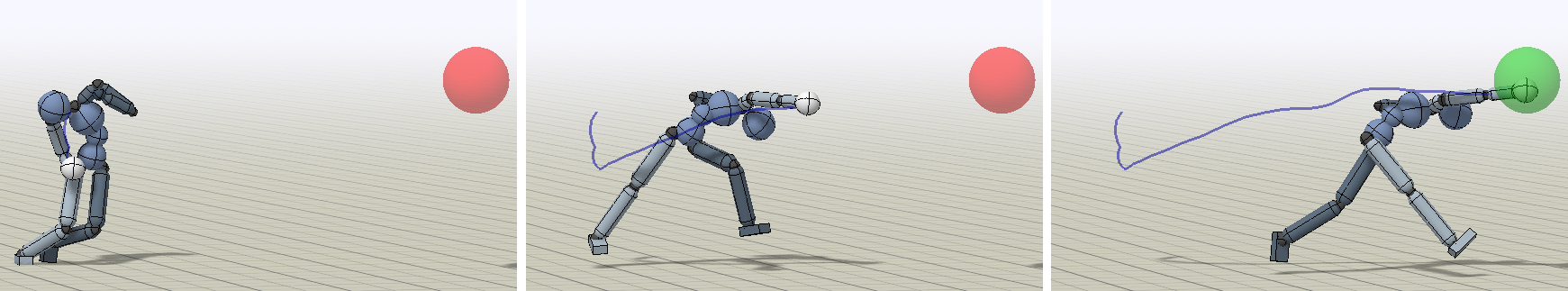}
    \vspace{-2em}
\caption{Policy trained for the throw task without a reference motion. Instead of throwing the ball, the character learns to run towards the target.}
\label{fig:throwNoRef}
\vspace{-0.1cm}
\end{figure}

In addition to imitating reference motions, the policies can also adapt the motions as needed to satisfy additional task objectives, such as following a target heading and throwing a ball to a randomly placed target. Performance statistics for each task are available in Table~\ref{tab:taskPerf}. To investigate the extent to which the motions are adapted for a particular task, we compared the performance of policies trained to optimize both the imitation objective $r^I$ and the task objective $r^G$ to policies trained only with the imitation objective. Table~\ref{tab:taskRewardPerf} summarizes the success rates of the different policies. For the throwing task, the policy trained with both objectives is able to hit the target with a success rate of $75\%$, while the policy trained only to imitate the baseball pitch motion is successful only in $5\%$ of the trials. Similarly, for the strike task, the policy trained with both objectives successfully hits $99\%$ of the targets, while the policy trained only to imitate the reference motion has a success rate of $19\%$. These results suggest that simply imitating the reference motions is not sufficient to successfully perform the tasks. The policies trained with the task objective are able to deviate from the original reference motion and developing additional strategies to satisfy their respective goals. We further trained policies to optimize only the task objective, without imitating a reference motion. The resulting policies are able to fulfill the task objectives, but without a reference motion, the policies develop unnatural behaviors, similar to those produced by prior deep RL methods. For the throw task, instead of throwing the ball, the policy adopts an awkward but functional strategy of running towards the target with the ball. Figure~\ref{fig:throwNoRef} illustrates this behavior.

\begin{table}[t]
{ \centering  
\caption{Success rate of policies trained with the imitation or task objectives disabled. Each policy is evaluated over 100 trials. Simply imitating the reference motions proves insufficient for fulfilling the task objectives. Training without a reference motion produces policies that develop awkward, but functional, strategies for satisfying the task objectives.}
\label{tab:taskRewardPerf}
\vspace{-0.25cm}
\begin{tabular}{|l|c|c|c|}
\hline
{\bf Environment} & {$\boldsymbol r^I + r^G$} & {$\boldsymbol r^I$} & {$\boldsymbol r^G$} \\ \hline
Humanoid: Strike - Spinkick & 99\% & 19\% & 55\% \\ \hline
Humanoid: Baseball Pitch - Throw & 75\% & 5\% & 93\% \\ \hline
\end{tabular} \\
}
\vspace{-0.5cm}
\end{table}

\subsection{Multi-Skill Integration}

Our method can construct policies for reproducing a wide range of individual motion clips. However, many more complex skills require choosing from among a set of potential behaviors. In this section, we discuss several methods by which our approach can be extended to combine multiple clips into a single compound skill.

\begin{figure}[t]
	\flushleft
     \subfigure{\includegraphics[width=1\columnwidth]{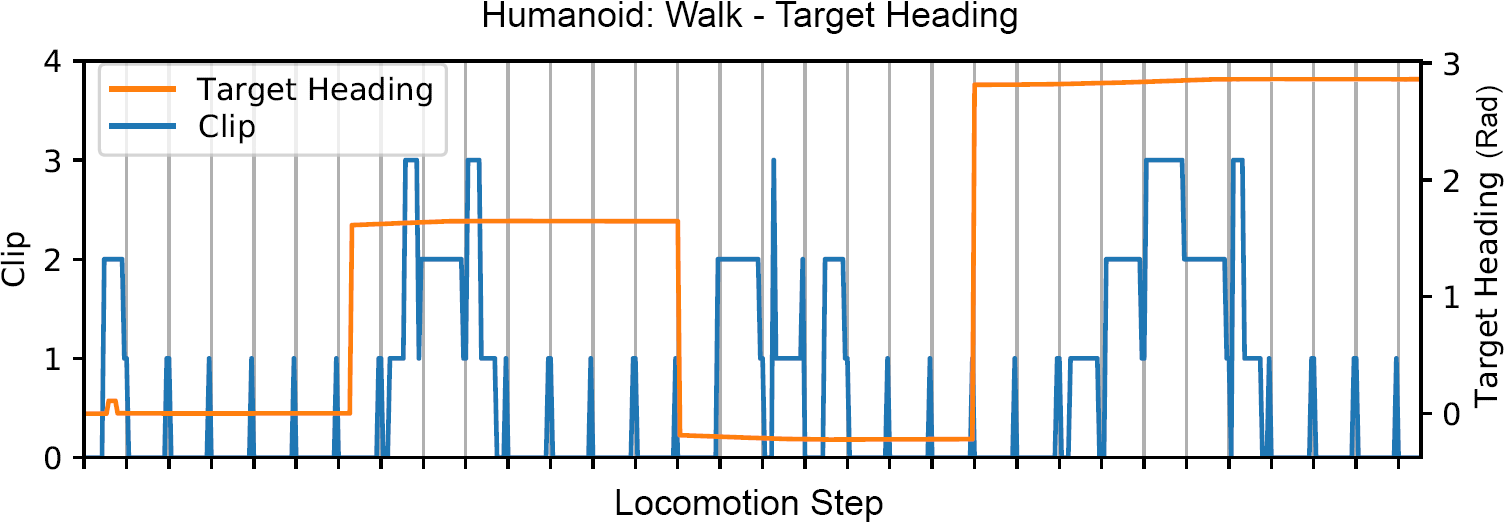}}
\vspace{-0.5cm}
\caption{Visualization of the walking reference motion clip that most closely matches the simulated character's motion as the character follows a changing target heading. Clip 0 is a forward walking motion, and clip 1-4 are turning motions. The policy is trained with the multi-clip reward.}
\label{fig:clipMatch}
\vspace{-0.5cm}
\end{figure}

\paragraph{Multi-Clip Reward:} To evaluate the multi-clip reward, we constructed an imitation objective from 5 different walking and turning clips.
Using this reward, a humanoid policy is then trained to walk while following a desired heading. The resulting policy learns to utilize a variety of agile stepping behaviours in order to follow the target heading. To determine if the policy is indeed learning to imitate multiple clips, we recorded the ID of the clip that best matches the motion of the simulated character at each timestep as it follows a changing target heading. The best matching clip is designated as the clip that yields the maximum reward at the current timestep. Figure~\ref{fig:clipMatch} records the best matching clips and the target heading over a 20$s$ episode. Clip 0 corresponds to a forward walking motion, and clips 1-4 are different turning motions. When the target heading is constant, the character's motion primarily matches the forward walk. When the target heading changes, the character's motion becomes more closely matched with the turning motions. After the character has realigned with the target heading, the forward walking clip once again becomes the best match. This suggests that the multi-clip reward function does indeed enable the character to learn from multiple clips of different walking motions. When using the multi-clip reward, we found that the clips should generally be from a similar type of motion (e.g. different kinds of walks and turns). Mixing very different motions, like a sideflip and a frontflip together, can result in the policy imitating only a subset of the clips. For more diverse clips, we found the composite policy to be a more effective integration method.

\paragraph{Skill Selector:} Using the one-hot vector representation, we trained a policy to perform various flips, and another to jump in different directions. The flip policy is trained to perform a frontflip, backflip, left sideflip, and right sideflip. The jump policy is trained to perform a forward, backward, left, and right jump. The first and last frame of each clip are duplicated to ensure that all clips within each set have the same cycle period. Once trained, the policies are able to perform arbitrary sequences of skills from their respective repertoires. The one-hot vector also provides a convenient interface for users to direct the character in real-time. Footage of the humanoid executing sequences of skills specified by a user is available in the supplemental video.

\paragraph{Composite Policy:} While the multi-clip objective can be effective for integrating multiple clips that correspond to the same category (e.g., walking), we found it to be less effective for integrating clips of more diverse skills. To integrate a more diverse corpus of skills, we constructed a composite policy from the backflip, frontflip, sideflip, cartwheel, spinkick, roll, getup face-down, and getup face-up policies.
The output of the value functions are normalized to be between [-1, 1] and the temperature is set to $\mathcal{T} = 0.3$. A new skill is sampled from the composite policy at the start of each cycle, and the selected skill is executed for a full cycle before selecting a new skill. To prevent the character from repeatedly executing the same skill, the policy is restricted to never sample the same skill in consecutive cycles. Unlike the skill selector policies, the individual policies in the composite policy are never explicitly trained to transition between different skills. By using the value functions to guide the selection of skills, the character is nonetheless able to successfully transition between the various skills. When the character falls, the composite policy activates the appropriate getup policy without requiring any manual scripting, as shown in the supplemental video.

\begin{figure}[t]
	\centering
    \subfigure{\includegraphics[width=0.49\columnwidth]{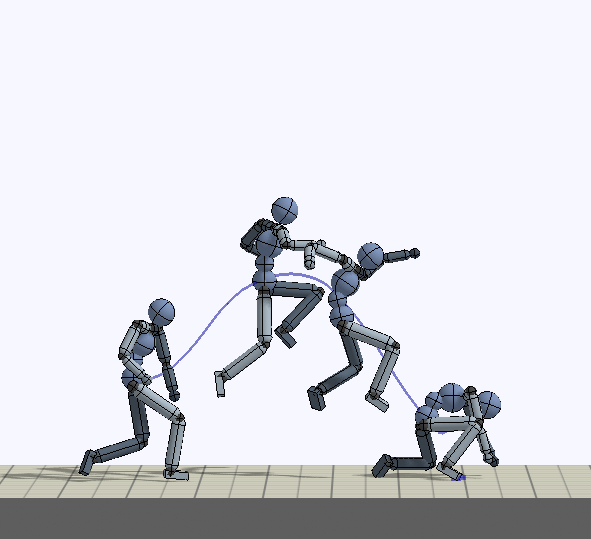}}
    \subfigure{\includegraphics[width=0.49\columnwidth]{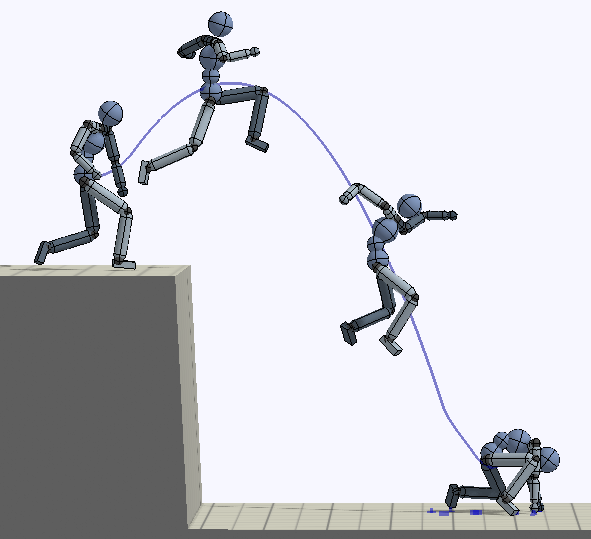}}
    \vspace{-1em}
\caption{\textbf{Left:} Original landing motion on flat terrain. \textbf{Right:} Policy trained to imitating landing motion while jumping down from a 2$m$ ledge. Despite being provided only with a motion recorded on flat terrain, the policy is able to adapt the skill to jump down from a tall ledge.}
\label{fig:landing}
\vspace{-0.6cm}
\end{figure}

\subsection{Retargeting} Due to modeling discrepancies between simulation and the real world, the dynamics under which a motion capture clip was recorded can differ dramatically from the dynamics of the simulated environments. Furthermore, keyframed motions may not be physically correct at all. To evaluate our framework's robustness to these discrepancies, we trained policies to perform similar skills with different character models, environments, and physics. 

\paragraph{Character Retargeting:} To demonstrate the system's capabilities in retargeting motions to different characters, we trained policies for walking, running, backflips and spinkicks on a simulated model of the Atlas robot from \url{http://www.mujoco.org/forum/index.php?resources/atlas-v5.16/}. The Atlas has a total mass of 169.8$kg$, almost four times the mass of the humanoid, as well as a different mass distribution. The serial 1D revolute joints in the original Atlas model are aggregated into 3D spherical joints to facilitate simpler retargeting of the reference motions. To retarget the motion clips, we simply copied the local joint rotations from the humanoid to the Atlas, without any further modification.
New policies are then trained for the Atlas to imitate the retargeted clips. Despite the starkly different character morphologies, our system is able to train policies that successfully reproduce the various skills with the Atlas model. Performance statistics of the Atlas policies are available in Table \ref{tab:perf}. The performance achieved by the Atlas policies are comparable to those achieved by the humanoid. Qualitatively, the motions generated by the Atlas character closely resemble the reference clips. To further highlight the differences in the dynamics of the characters, we evaluated the performance of directly applying policies trained for the humanoid to the Atlas. The humanoid policies, when applied to the Atlas, fail to reproduce any of the skills, achieving a normalized return of 0.013 and 0.014 for the run and backflip, compared to 0.846 and 0.630 achieved by policies that were trained specifically for the Atlas but using the same reference clips.

\paragraph{Environment Retargeting:} While most of the reference motions were recorded on flat terrain, we show that the policies can be trained to adapt the motions to irregular environments. First, we consider the case of retargeting a landing motion, where the original reference motion is of a human subject jumping and landing on flat ground. From this reference motion, we trained a character to reproduce the motion while jumping down from a 2$m$ ledge. Figure~\ref{fig:landing} illustrates the motion from the final policy. The system was able to adapt the reference motion to a new environment that is significantly different from that of the original clip.

\begin{figure}[t]
	\flushleft
    \subfigure{\includegraphics[width=0.49\columnwidth]{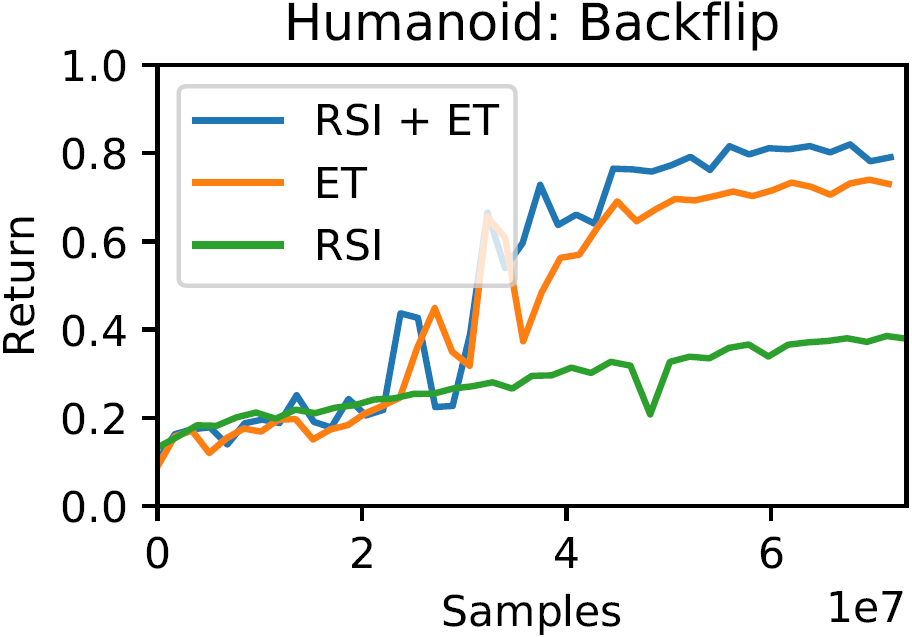}}
     \subfigure{\includegraphics[width=0.49\columnwidth]{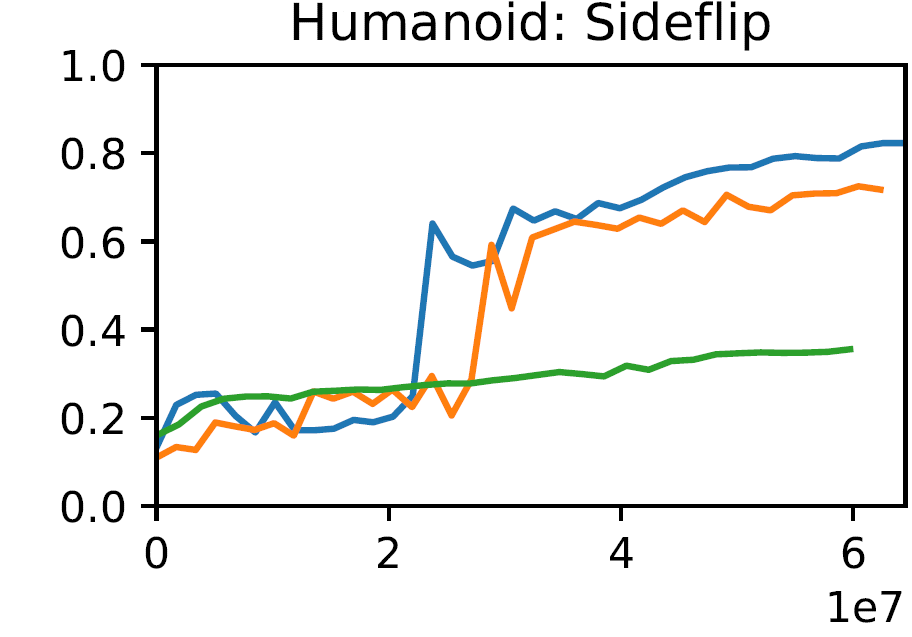}}\\
     \vspace{-0.5em}
     \subfigure{\includegraphics[width=0.49\columnwidth]{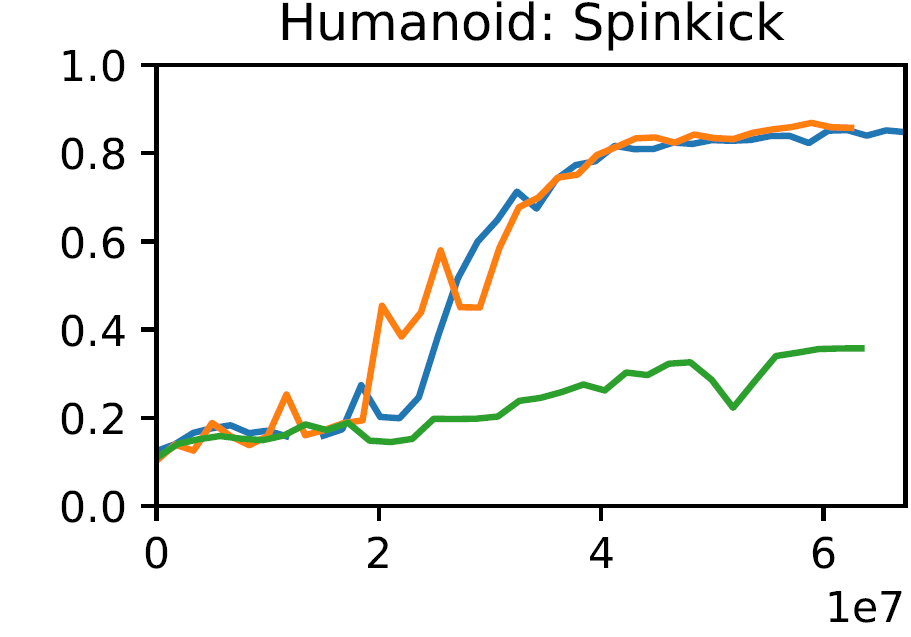}}
     \subfigure{\includegraphics[width=0.49\columnwidth]{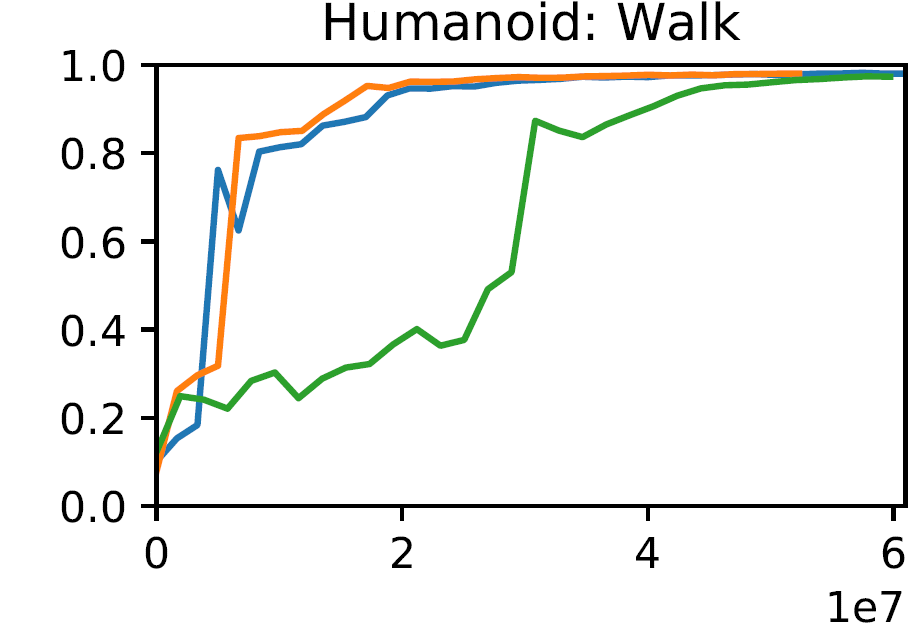}}\\
     \vspace{-1.5em}
\caption{Learning curves for policies trained with and without reference state initialization (RSI) and early termination (ET).}
\label{fig:curvesRSIET}
\vspace{-0.2cm}
\end{figure}

\begin{table}[t]
{ \centering  
\caption{Normalized average returns of policies trained with and without reference state initialization (RSI) and early termination (ET).}
\label{tab:compRSIET}
\vspace{-0.25cm}
\begin{tabular}{|l|c|c|c|}
\hline
{\bf Skill} & {\bf RSI + ET} & {\bf ET} & {\bf RSI} \\ \hline
Backflip &  0.791 & 0.730 & 0.379 \\ \hline
Sideflip & 0.823 & 0.717 & 0.355 \\ \hline
Spinkick &  0.848 & 0.858 & 0.358 \\ \hline
Walk & 0.980 & 0.981 & 0.974 \\ \hline
\end{tabular} \\
}
\vspace{-0.5cm}
\end{table}

Next, we explore vision-based locomotion in more complex procedurally generated environments. By augmenting the networks with a heightmap input, we are able to train the humanoid to run across terrains consisting of random obstacles. Examples of the environments are available in Figure~\ref{fig:terrains}. Over the course of training, the policies are able to adapt a single clip of a forward running motion into a variety of strategies for traversing across the different classes of obstacles. Furthermore, by training a policy to imitate a balance beam walk, the character learns to follow a narrow winding path using only a heightmap for pathfinding. The balance beam policy was trained with only a forward walking clip, but is nonetheless able to develop turning motions to follow the winding path. In addition to the humanoid, we also trained a policy for the Atlas to climb up stairs with irregular step heights. The policy was able to adapt the original walking clip on flat terrain to climb the steps, although the resulting motion still exhibits an awkward gait. We suspect that the problem is partly related to the walking reference motion being ill-suited for the stairs environment.

\vspace{-0.1cm}
\paragraph{Physics Retargeting:} To further evaluate the framework's robustness to discrepancies between the dynamics of the motion capture data and simulation, we trained policies to perform a spinkick and cartwheel under moon gravity ($1.622 m/s^2$). Despite the difference in gravity, the policies were able to adapt the motions to the new dynamics, achieving a return of 0.792 for the spinkick and 0.688 for the cartwheel.

\vspace{-0.1cm}
\subsection{Ablations} To evaluate the impact of our design decisions, we compare our full method against alternative training schemes that disable some of the components. We found that the reference state initialization and early termination are two of the most important components of our training procedure. The comparisons include training with reference state initialization and with a fixed initial state, as well as training with early termination and without early termination. Without early termination, each episode is simulated for the full 20$s$. Figure~\ref{fig:curvesRSIET} compares the learning curves with the different configurations and Table~\ref{tab:compRSIET} summarizes the performance of the final policies. During evaluation, early termination is disabled and the character is initialized to a fixed state at the start of each episode. Due to the time needed to train each policy, the majority of performance statistics are collected from one run of the training process. However, we have observed that the behaviors are consistent across multiple runs. Early termination proves to be crucial for reproducing many of the skills. By heavily penalizing the character for making undesirable contacts with the ground, early termination helps to eliminate local optima, such as those where the character falls and mimes the motions as it lies on the ground. RSI also appears vital for more dynamic skills that have significant flight phases, such as the backflip. While the policies trained without RSI appear to achieve a similar return to those trained with RSI, an inspection of the resulting motions show that, without RSI, the character often fails to reproduce the desired behaviours. For the backflip, without RSI, the policy never learns to perform a full mid-air flip. Instead, it performs a small backwards hop while remaining upright. 

\vspace{-0.1cm}
\subsection{Robustness}
To determine the policies' robustness to external perturbations, we subjected the trained policies to perturbation forces and recorded the maximum force the character can tolerate before falling. The perturbation forces are applied halfway through a motion cycle to the character's pelvis for 0.2$s$. The magnitude of the force is increased by 10$N$ until the character falls. This procedure is repeated for forces applied along the forward direction (x-axis) and sideway direction (z-axis). Table~\ref{tab:perturbTest} summarizes the results from the experiments on the humanoid. The learned policies show comparable-or-better robustness than figures reported for SAMCON \cite{2016-TOG-controlGraphs}. The run policy is able to recover from $720N \times 0.2s$ forward pushes, while the spin-kick policy is able to survive $600N \times 0.2s$ perturbations in both directions. Note that no external perturbations are applied during the training process; we suspect that the policies' robustness is in large part due to the exploration noise applied by the stochastic policy used during training.

\begin{table}[t]
{ \centering  
\caption{Maximum forwards and sideways push each policy can tolerate before falling. Each push is applied to the character's pelvis for 0.2$s$.}
\label{tab:perturbTest}
\vspace{-0.25cm}
\begin{tabular}{|l|c|c|c|}
\hline
{\bf Skill} & {\bf Forward (N)} & {\bf Sideway (N)} \\ \hline
Backflip & 440 & 100 \\ \hline
Cartwheel & 200 & 470 \\ \hline
Run & 720 & 300 \\ \hline
Spinkick & 690 & 600 \\ \hline
Walk & 240 & 300 \\ \hline
\end{tabular} \\
}
\vspace{-0.5cm}
\end{table}

\section{Discussion and Limitations}

We presented a data-driven deep reinforcement learning framework for training control policies for simulated characters. 
We show that our method can produce a broad range of challenging skills. The resulting policies are highly robust and produce natural motions that are nearly indistinguishable from the original motion capture data in the absence of perturbations. Our framework is able to retarget skills to a variety of characters, environments, and tasks, and multiple policies can be combined into composite policies capable of executing multiple skills.

Although our experiments illustrate the flexibility of this approach, there are still numerous limitations to be addressed in future work. First, our policies require a phase variable to be synchronized with the reference motion, which advances linearly with time. This limits the ability of the policy to adjust the timing of the motion, and lifting this limitation could produce more natural and flexible perturbation recoveries. Our multi-clip integration approach works well for small numbers of clips, but has not yet been demonstrated on large motion libraries. The PD controllers used as the low-level servos for the characters still require some insight to set properly for each individual character morphology. The learning process itself is also quite time consuming, often requiring several days per skill, and is performed independently for each policy. Although we use the same imitation reward across all motions, this is still currently based on a manually defined state-similarity metric.
The weighting of the reward terms also needs to be defined with some care.

We believe this work nevertheless opens many exciting directions for exploration.
In future work, we wish to understand how the policies might be deployed on robotic systems, 
as applied to locomotion, dexterous manipulation, and other tasks.
It would be interesting to understand the learned control strategies and compare them to the equivalent human strategies.
We wish to integrate diverse skills that would enable a character to perform more challenging tasks and more complex 
interactions with their environments. Incorporating hierarchical structure is likely to be beneficial towards this goal. 

\section*{Acknowledgements}
We thank KangKang Yin, Libin Liu, and Ziva Dynamics for providing reference motion data for this project, the anonymous reviewers for their helpful feedback, and AWS for providing computational resources. This research was funded by an NSERC Postgraduate Scholarship, a Berkeley Fellowship for Graduate Study, and ONR PECASE N000141612723.

\bibliographystyle{ACM-Reference-Format}
\bibliography{ImitateRL}

\newpage

\section*{Supplementary Material}

\appendix

\section{Multi-step Returns}
We will refer to the return $R_t = \sum_{l = 0}^{T - t} \gamma^l r_{t + l}$ as the Monte-Carlo return. $R_t$ provides an unbiased sample of the expected return at a given state, but due to stochasticity from the dynamics of the environment and policy, each reward $r_t$ can be a random variable, the sum of which can result in a high variance estimator of the expected return. Alternatively, an $n$-step return can be used to provide a lower-variance estimator at the cost of introducing some bias. The $n$-step return can be computed by truncating the sum of returns after $n$ steps, and approximating the return from the remaining steps using a value function $V(s)$:
\[R^{(n)}_t = \sum_{l = 0}^{n-1} \gamma^{l} r_{t + l} + \gamma^n V(s_{t + n}).\]
$R^{(1)}_t = r_t + \gamma V(s_{t + 1})$ results in the 1-step return commonly used in $Q$-learning \cite{mnih2015humanlevel} and $R^{(\infty)}_t = \sum_{l = 0}^{T - t} \gamma^{t} r_{t + l} = R_t$ recovers the original Monte-Carlo return, given that all rewards after the horizon $T$ are 0. While $R^{(\infty)}_t$ provides an unbiased but high variance estimator, $R^{(1)}_t$ provides a biased but low variance estimator. Therefore, $n$ acts as a trade-off between bias and variance for the value estimator.

Another method to trade off between the bias and variance of the estimator is to use a $\lambda$-return \cite{Sutton1998}, calculated as an exponentially-weighted average of $n$-step returns with decay parameter $\lambda$:
\[R_t(\lambda) = (1 - \lambda) \sum_{n=1}^\infty \lambda^{n - 1} R_t^n.\]
Assuming all rewards after step $T$ are 0, such that $R_t^n = R_t^{T - t}$ for all $n \geq T - t$, the infinite sum can be calculated according to
\[R_t(\lambda) = (1 - \lambda) \sum_{n=1}^{T - t - 1} \lambda^{n - 1} R_t^n + \lambda^{T - t - 1} R_t^{T - t}\]
$\lambda = 0$ reduces to the single-step return $R_t^{(1)}$, and $\lambda = 1$ recovers the Monte-Carlo return $R_t$. Intermediate values of $\lambda \in (0, 1)$ produces interpolants that can be used to balance the bias and variance of the value estimator. 

Updating the value function using the temporal difference computed with the $\lambda$-return results in the TD($\lambda$) algorithm \cite{Sutton1998}. Similarly, estimating the advantage with the $\lambda$-return, yields the generalized advantage estimator GAE($\lambda$) \cite{SchulmanMLJA15}:
\[\hat{\mathcal{A}}_t = R_t(\lambda) - V(s_t).\]
Empirically, the $\lambda$-return has been shown to produce better performance than simply using an $n$-step return \cite{Sutton1998,SchulmanMLJA15}.

\section{Off-policy Learning}
In its previously stated form, the gradient of the expected return $\triangledown_\theta J(\theta)$ is estimated with respect to the current policy parameters $\theta$. As such, data collected from the current policy can be justified for use only in performing a single gradient step, after which a new batch of data is required to estimate the gradient with respect to the updated parameters. Data efficiency can be improved by introducing importance sampling, which provides an unbiased estimate of the policy gradient using only off-policy samples from an older policy $\pi_{\theta_{old}}$:
\[\triangledown_\theta J(\theta) = \mathbb{E}_{s_t \sim d_{\theta_{old}}(s_t), a_t \sim \pi_{\theta_{old}}(a_t | s_t)} \left[ w_t(\theta) \triangledown_\theta \mathrm{log}(\pi_\theta(a_t | s_t)) \mathcal{A}_t \right] \]
\[w_t(\theta) = \frac{\pi_\theta(a_t | s_t)}{\pi_{\theta_{old}}(a_t | s_t)}\]
The importance-sampled policy gradient can be interpreted as optimizing the surrogate objective
\[L^{IS}(\theta) = \mathbb{E}_{s_t \sim d_{\theta_{old}}(s_t), a_t \sim \pi_{\theta_{old}}(a_t | s_t)} \left[ w_t(\theta)\mathcal{A}_t \right]. \]
With $L^{IS}$, the same batch of data can be used to perform multiple update steps for $\theta$. While in theory importance sampling allows for the policy gradient to be estimated with data collected from any policy with sufficient support, we will consider only the case where the data is collected from a previous set of policy parameters.

\section{Proximal Policy Optimization}
In practice, the policy gradient estimator discussed thus far suffers from high variance, often leading to instability during learning, where the policy's performance fluctuates drastically between iterations. While problems due to noisy gradients can be mitigated by using large batches of data per update, policy gradient algorithms can still be extremely unstable. Trust region methods have been proposed as a technique for improving the stability of policy gradient algorithms \citep{SchulmanLMJA15}. Trust Region Policy Optimization (TRPO) optimizes the same objective $J(\theta)$ but includes an additional KL-divergence constraint to prevent the behaviour of the current policy $\pi_\theta$ from deviating too far from the previous policy $\pi_{\theta_{old}}$.

\begin{equation*}
\begin{aligned}
\underset{\theta}{\text{max}} 
& \qquad J(\theta) \\
\text{s.t.}
& \qquad \mathbb{E}_{s_t \sim d_\theta(s_t)} \left[ KL \left( \pi_{\theta_{old}}( \cdot | s_t) \middle| \pi_\theta( \cdot | s_t) \right) \right] \leq \delta_{KL}
\end{aligned}
\end{equation*}
\\
$\delta_{KL}$ is a hyper parameters that defines the trust region. TRPO has been successfully applied to solve a wide variety of challenging RL problems \citep{DuanCHSA16,GAIL2016,RajeswaranGLR16}. However, ensuring that the constraint is satisfied can be difficult, and it is often instead enforced approximately using the conjugate gradient algorithm with an adaptive stepsize selected by a line search to ensure the approximate constraint is satisfied \citep{SchulmanLMJA15}.

Proximal Policy Optimization (PPO) is a variant of TRPO, where the hard constraint is replaced by optimizing a surrogate loss \cite{PPO17}. In this work, we will be concerned mainly with the clipped surrogate loss $L^{CLIP}(\theta)$ defined according to
\[L^{CLIP}(\theta) = \mathbb{E}_{s_t, a_t} \left[ \mathrm{min} \left(w_t(\theta) \mathcal{A}_t, \mathrm{clip} \left(w_t(\theta), 1 - \epsilon, 1 + \epsilon \right)  \mathcal{A}_t \right) \right]. \]
\\
When $\theta = \theta_{old}$, $w_t(\theta) = 1$, but as $\theta$ is changed, the likelihood ratio moves away from 1. Therefore, the likelihood ratio can be interpreted as a measure of the similarity between two policies. To discourage the policies from deviating too far apart, $\mathrm{clip}\left( w_t(\theta), 1 - \epsilon, 1 + \epsilon \right)$ sets the gradient to zero whenever the ratio is more than $\epsilon$ away from 1. This term therefore serves a similar function to the KL-divergence constraint in TRPO. The minimum is then taken between the clipped and unclipped advantages to create a lower bound of $L(\theta)$.

\section{Learning Algorithm} 
Algorithm~\ref{alg:PPO} summarizes the common learning procedure used to train all policies. Policy updates are performed after a batch of $m = 4096$ samples has been collected. Minibatches of size $n = 256$ are then sampled from the data for each gradient step. A discount factor $\gamma = 0.95$ is used for all motions. $\lambda = 0.95$ is used for both TD($\lambda$) and GAE($\lambda$). The likelihood ratio clipping threshold is set to $\epsilon = 0.2$. A stepsize of $\alpha_v = 10^{-2}$ is used for the value function. A policy step size of $\alpha_\pi = 5 \times 10^{-5}$ is used for the humanoid and Atlas, and $\alpha_\pi = 2 \times 10^{-5}$ for the dragon and T-Rex. Once gradients have been computed, the network parameters are updated using stochastic gradient descent with momentum 0.9. The same hyperparameter settings are used for all characters and skills, with the exception of the step size. Humanoid policies for imitating individual skills typically require about 60 million samples to train, requiring about 2 days on an 8-core machine. All simulation and network updates are performed on the CPU and no GPU acceleration is used.

\begin{algorithm}[h!]
\caption{Proximal Policy Optimization}
\label{alg:PPO}
\begin{algorithmic}[1]
\STATE{$\theta \leftarrow$ random weights}
\STATE{$\psi \leftarrow$ random weights}
\WHILE{not done}
	\STATE{$s_0 \leftarrow$ sample initial state from reference motion}
    \STATE{Initialize character to state $s_0$}
	\FOR{step $= 1,...,m$}
		\STATE{$s \leftarrow$ start state}
		\STATE{$a \sim  \pi_\theta(a | s)$}
        
		\STATE Apply $a$ and simulate forward one step
		\STATE{$s' \leftarrow$ end state}
		\STATE{$r \leftarrow$ reward}
		\STATE{record $(s, a, r, s')$ into memory $D$}
	\ENDFOR

	\item[]
    \STATE{$\theta_{old} \leftarrow \theta$}
    \FOR{each update step}
    \STATE{Sample minibatch of $n$ samples $\{(s_i, a_i, r_i, s'_i)\}$ from $D$}
    
    \item[]
	\STATE{Update value function:}
	\FOR{each $(s_i, a_i, r_i, s'_i)$}
		\STATE{$y_i \leftarrow $ compute target values using TD($\lambda$)}
	\ENDFOR
    \STATE{$\psi \leftarrow \psi + \alpha_v \left( \frac{1}{n} \sum_{i} \triangledown_{\psi} V_\psi(s_{i}) (y_{i} - V(s_{i}))     \right)$}
	
    \item[]
	\STATE{Update policy:}
    \FOR{each $(s_i, a_i, r_i, s'_i)$}
		\STATE{$\mathcal{A}_i \leftarrow $ compute advantage using $V_\psi$ and GAE}
        \STATE{$w_i(\theta) \leftarrow \frac{\pi_\theta(a_i | s_i)}{\pi_{\theta_{old}}(a_i | s_i)}$}
	\ENDFOR

    \STATE{$\theta \leftarrow \theta + \alpha_\pi \frac{1}{n} \sum_{i} \triangledown_{\theta} \mathrm{min} \left( w_i(\theta) \mathcal{A}_i, \mathrm{clip} \left(w_i(\theta), 1 - \epsilon, 1 + \epsilon \right) \mathcal{A}_i \right)$}

\ENDFOR
        
\ENDWHILE
\end{algorithmic}
\end{algorithm}

\begin{figure*}[h]
	\flushleft
     	\subfigure{\includegraphics[width=0.5\columnwidth]{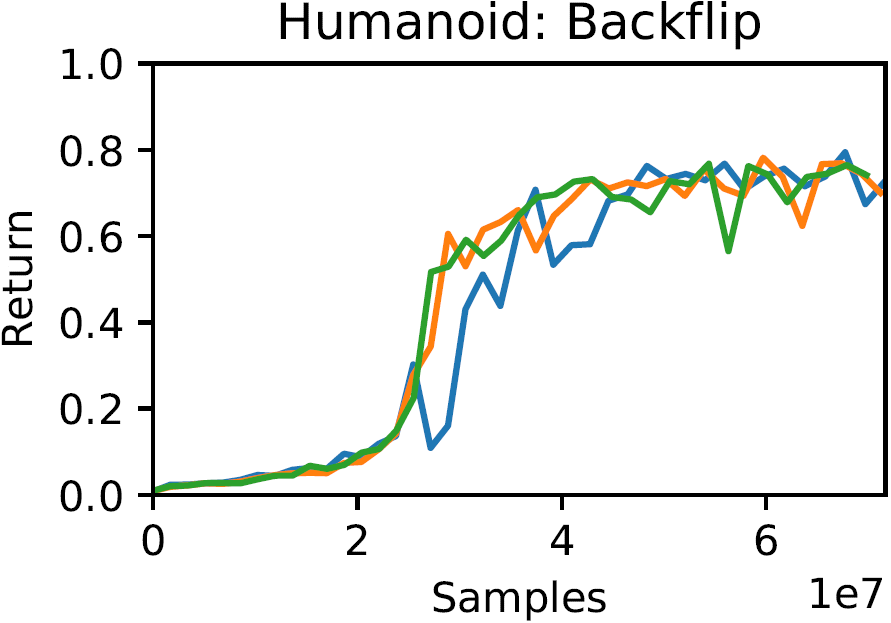}}
    	\subfigure{\includegraphics[width=0.5\columnwidth]{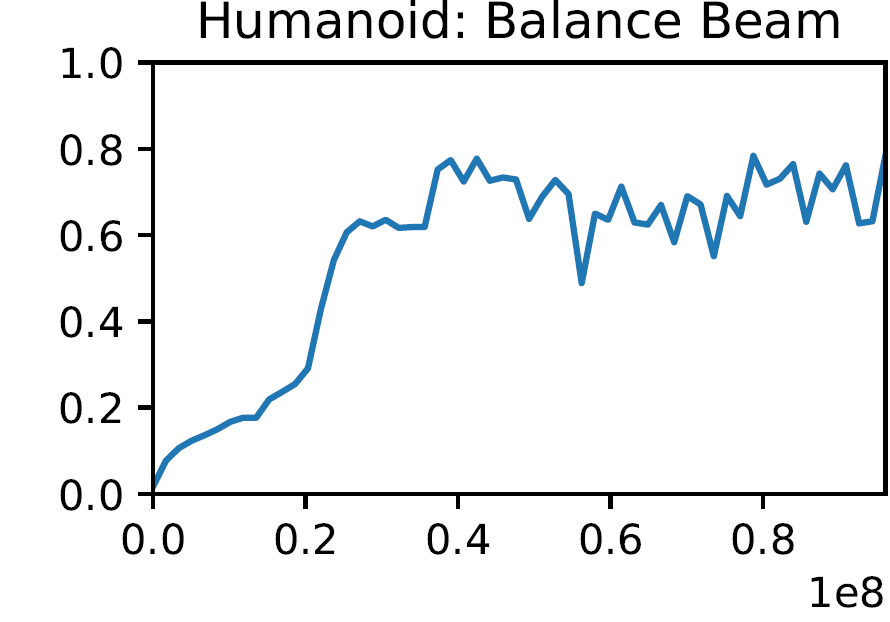}}
    	\subfigure{\includegraphics[width=0.5\columnwidth]{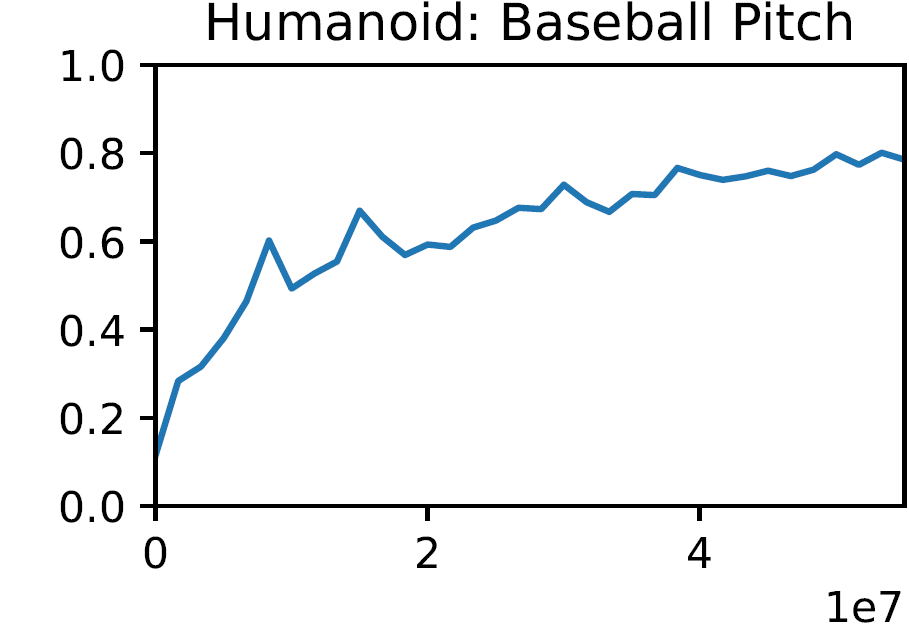}}
     	\subfigure{\includegraphics[width=0.5\columnwidth]{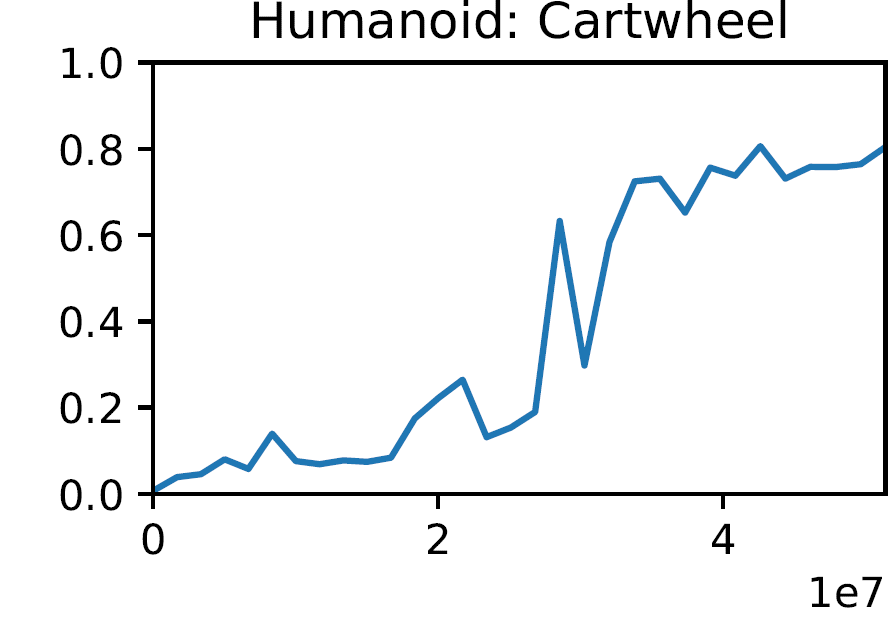}}
     	\subfigure{\includegraphics[width=0.5\columnwidth]{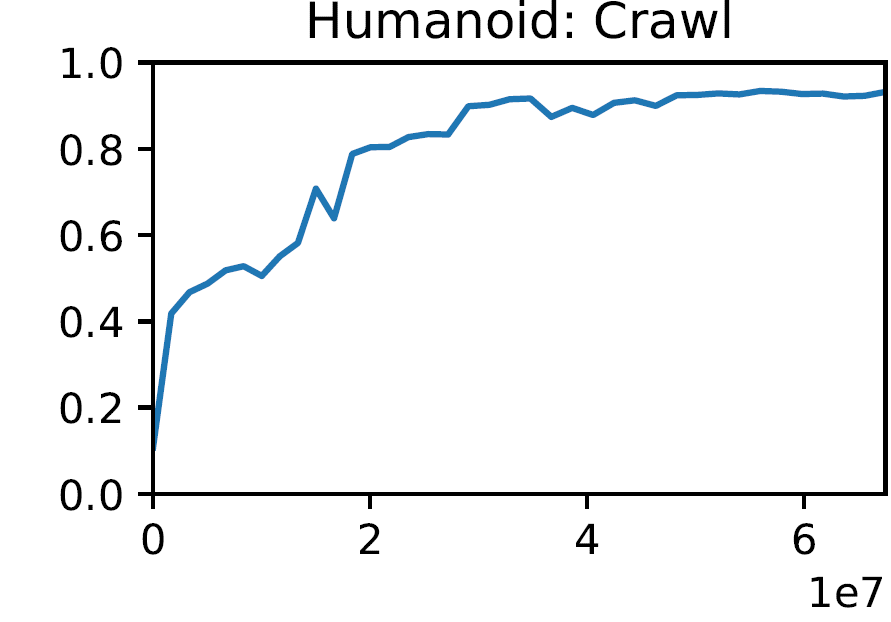}}
     	\subfigure{\includegraphics[width=0.5\columnwidth]{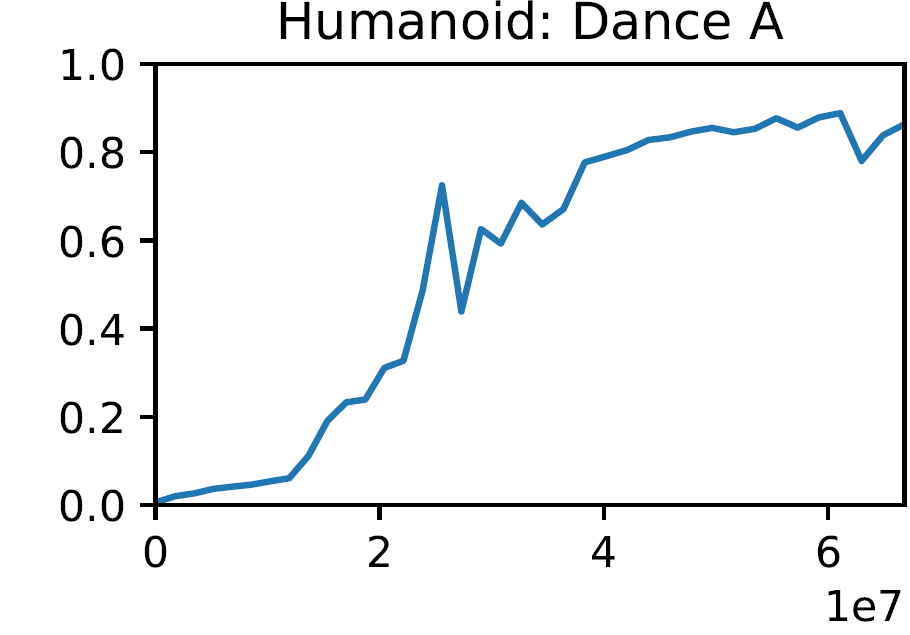}}
     	\subfigure{\includegraphics[width=0.5\columnwidth]{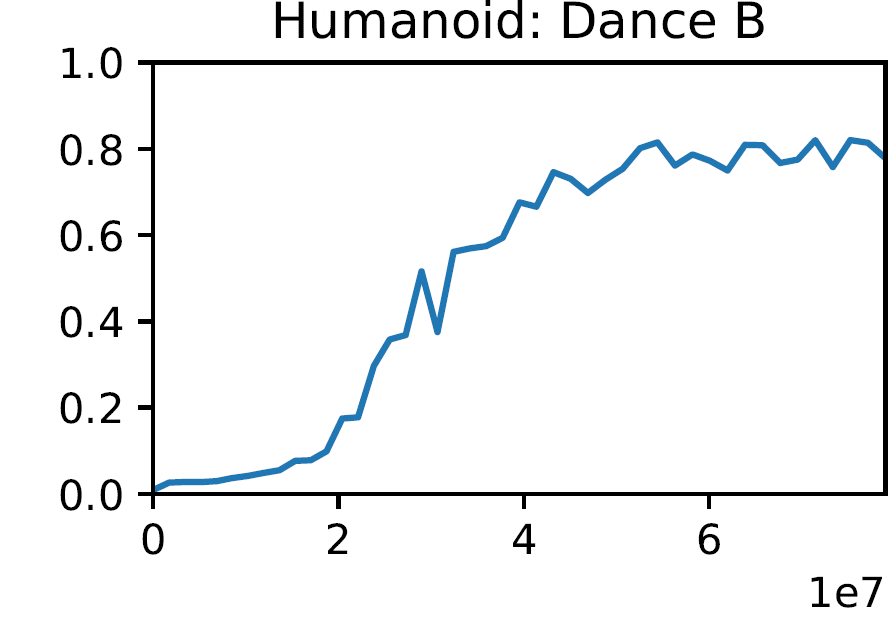}}
     	\subfigure{\includegraphics[width=0.5\columnwidth]{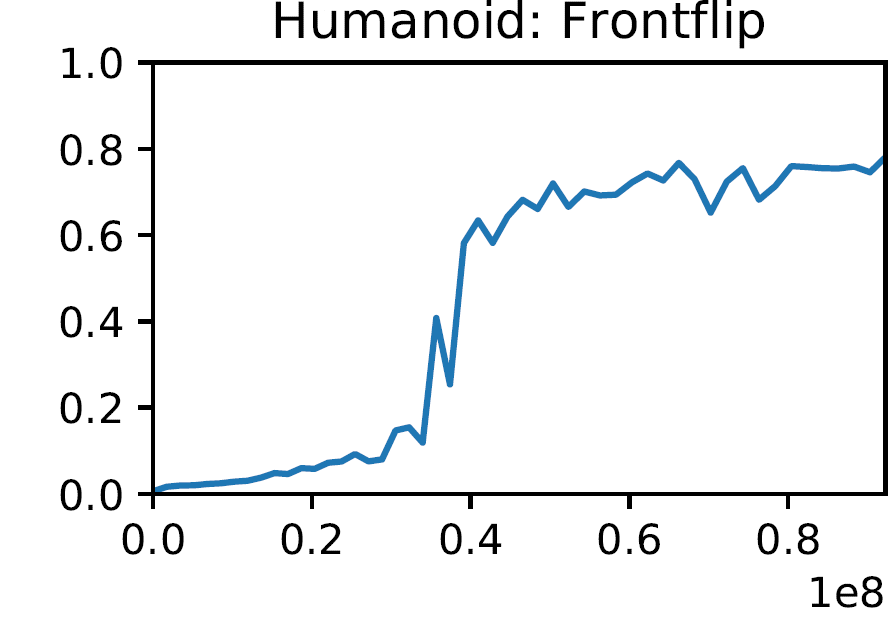}}
    	\subfigure{\includegraphics[width=0.5\columnwidth]{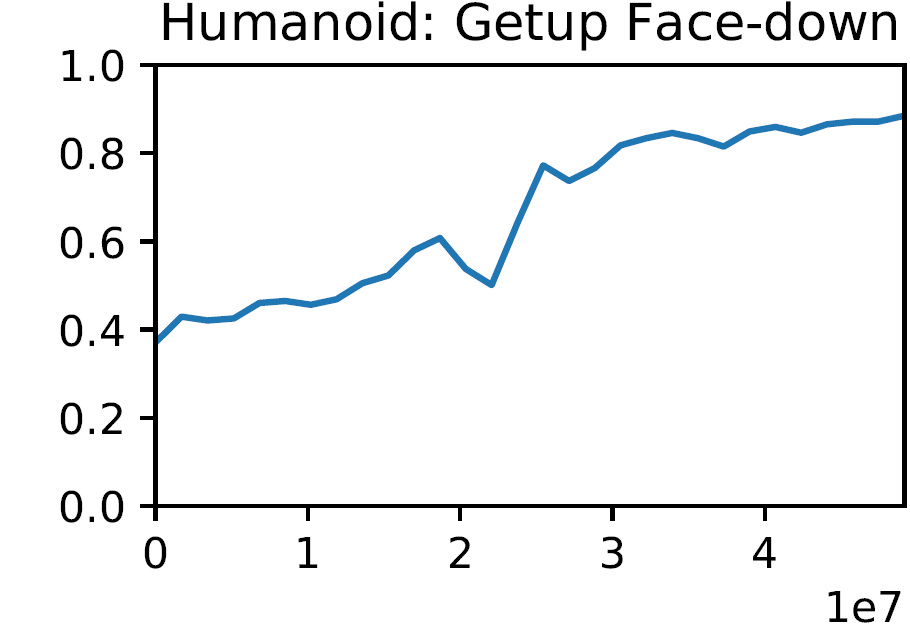}}
    	\subfigure{\includegraphics[width=0.5\columnwidth]{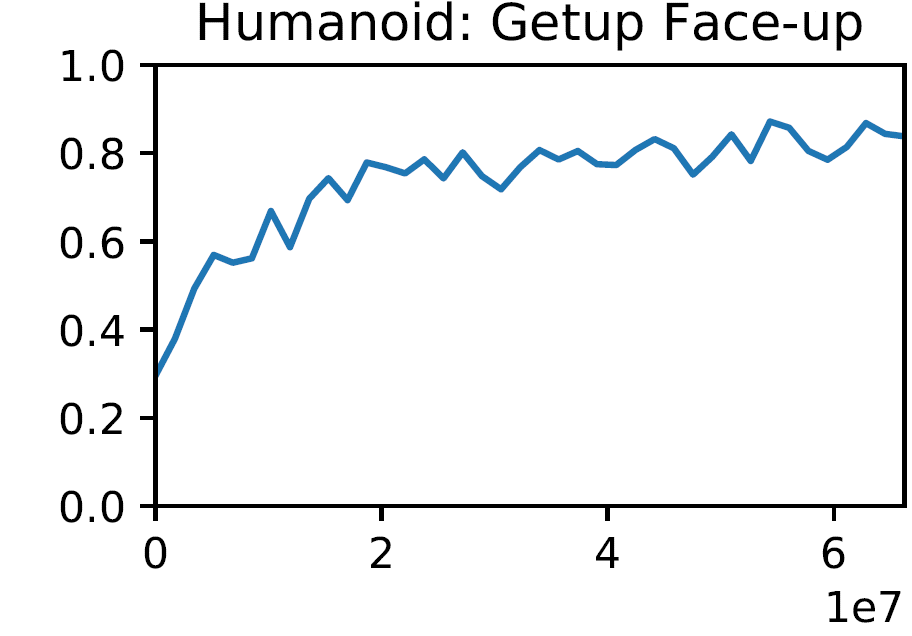}}
    	\subfigure{\includegraphics[width=0.5\columnwidth]{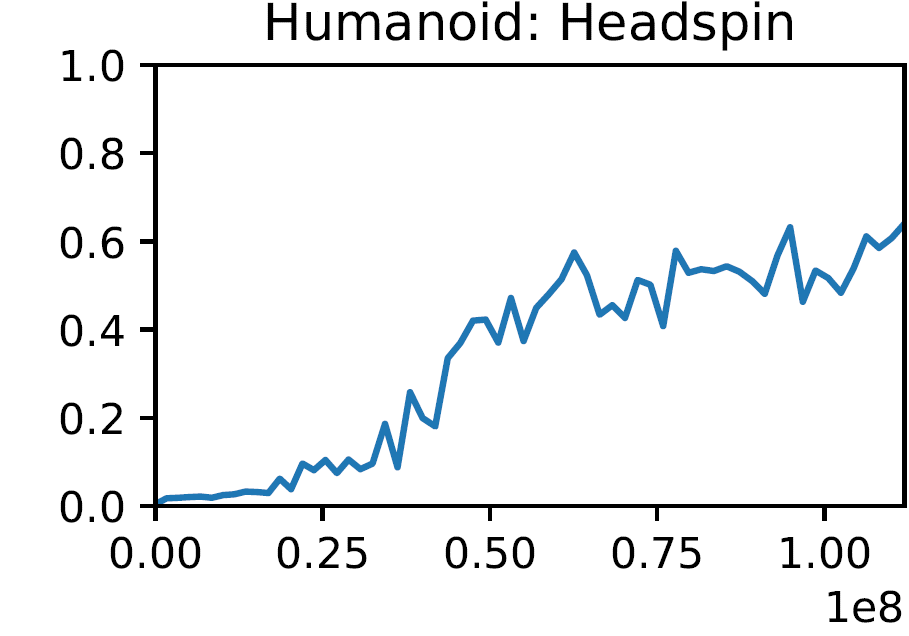}}
    	\subfigure{\includegraphics[width=0.5\columnwidth]{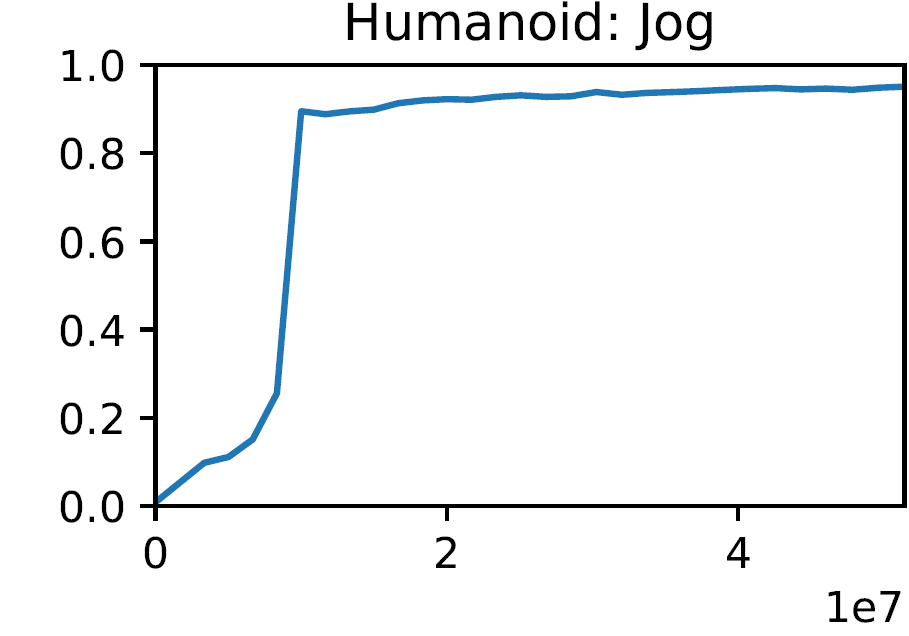}}
     	\subfigure{\includegraphics[width=0.5\columnwidth]{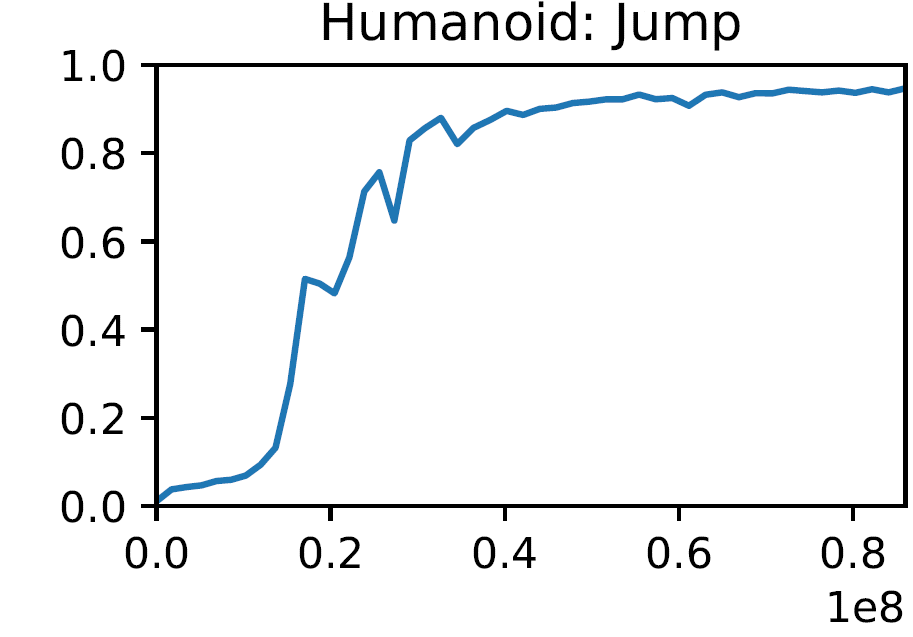}}
    	\subfigure{\includegraphics[width=0.5\columnwidth]{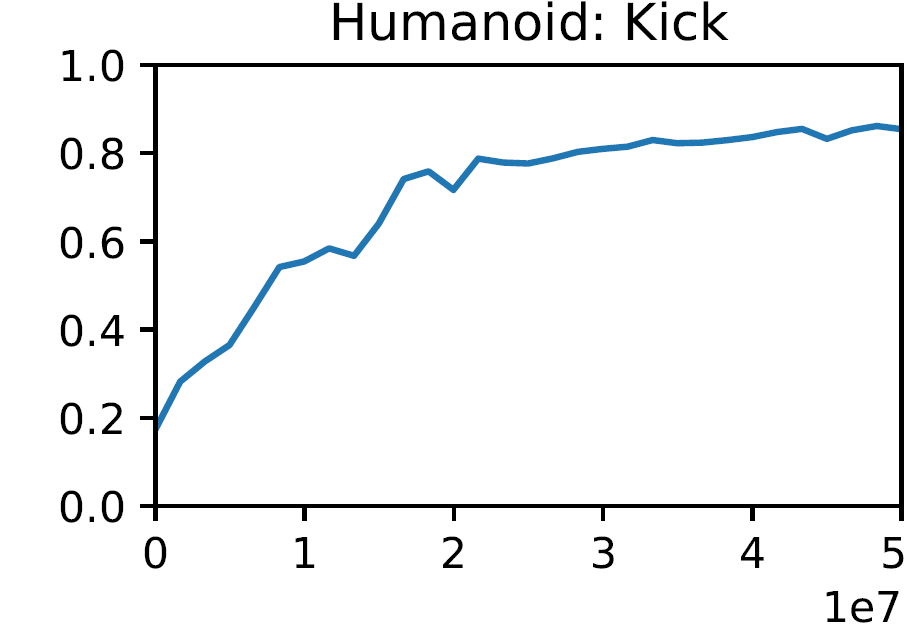}}
    	\subfigure{\includegraphics[width=0.5\columnwidth]{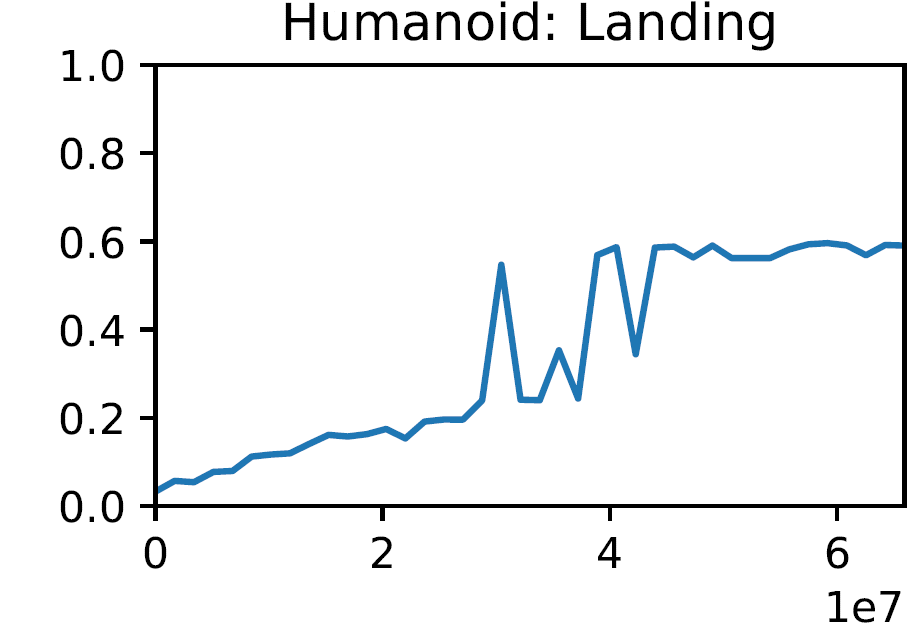}}
    	\subfigure{\includegraphics[width=0.5\columnwidth]{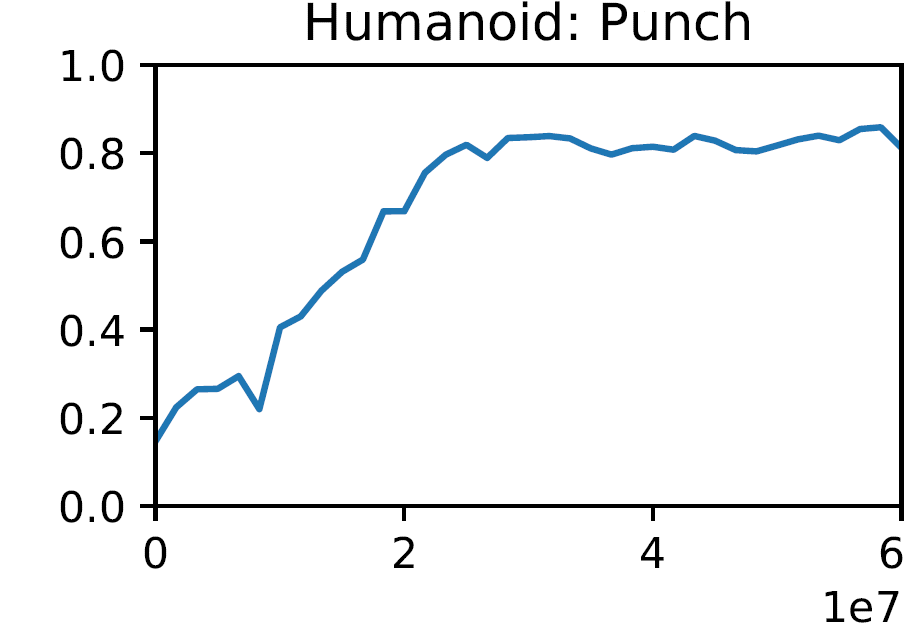}}
     	\subfigure{\includegraphics[width=0.5\columnwidth]{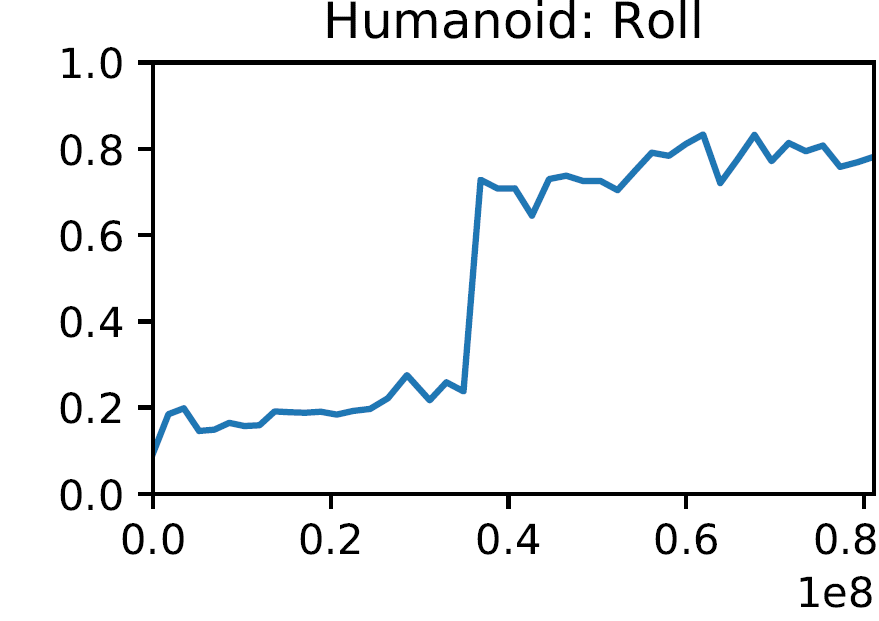}}
     	\subfigure{\includegraphics[width=0.5\columnwidth]{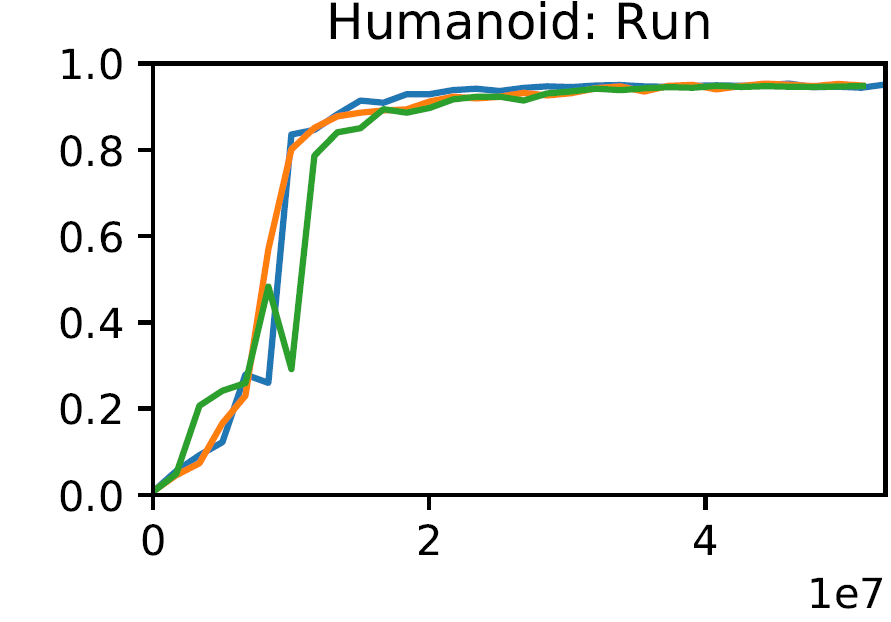}}
	 	\subfigure{\includegraphics[width=0.5\columnwidth]{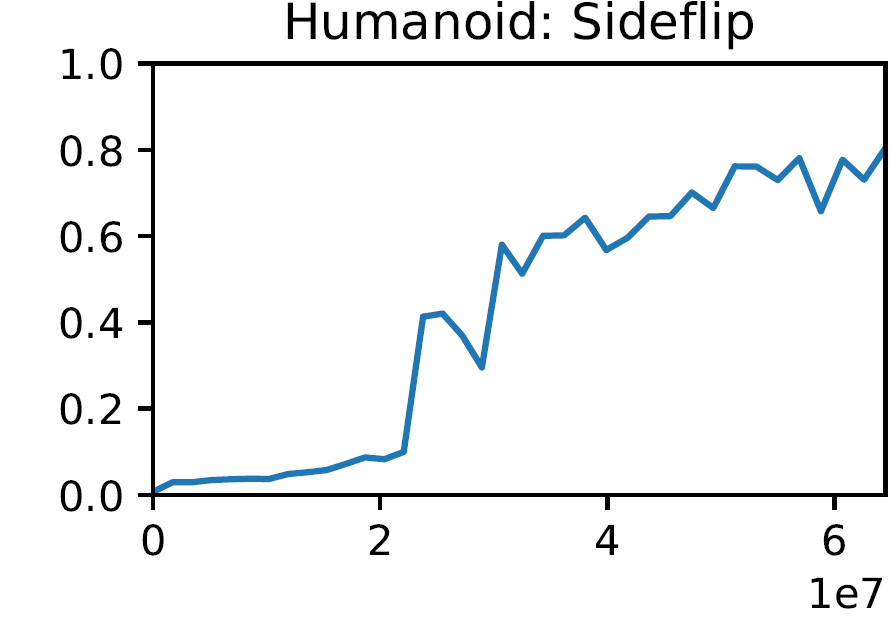}}
     	\subfigure{\includegraphics[width=0.5\columnwidth]{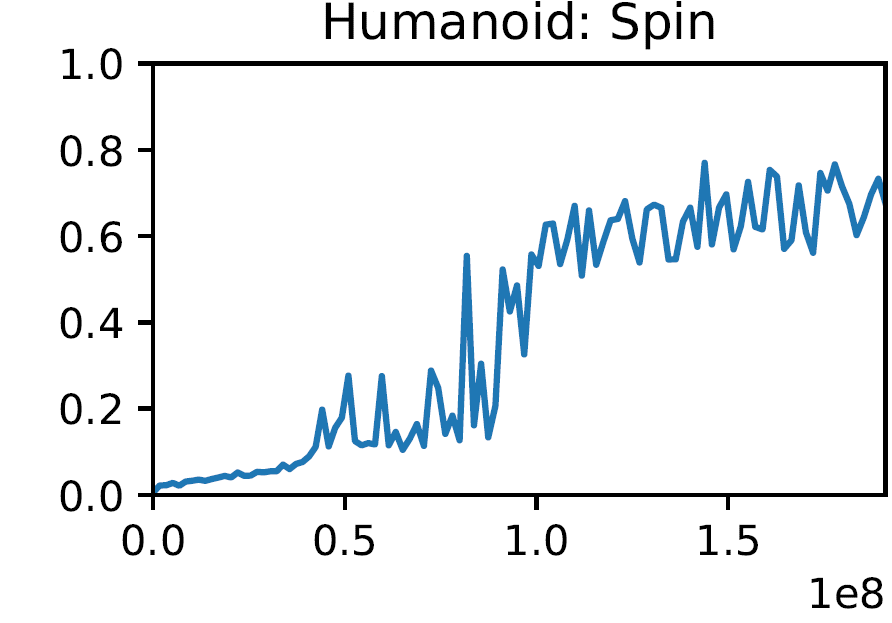}}
     	\subfigure{\includegraphics[width=0.5\columnwidth]{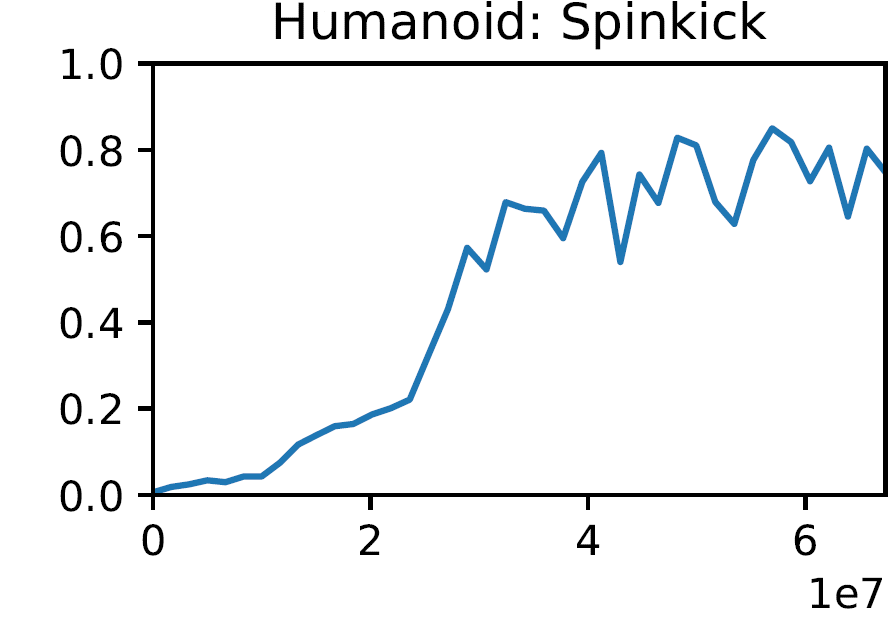}}
    	\subfigure{\includegraphics[width=0.5\columnwidth]{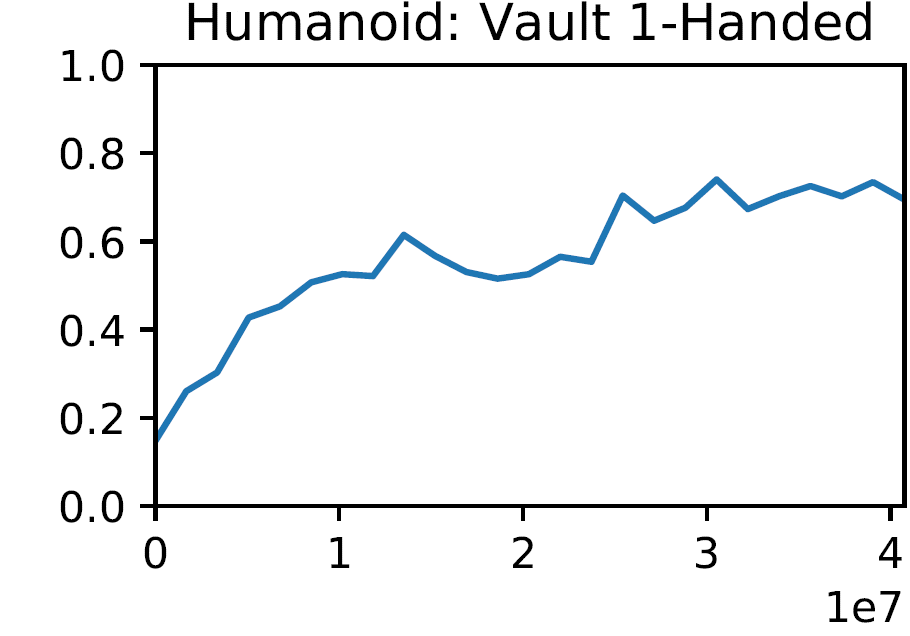}}
    	\subfigure{\includegraphics[width=0.5\columnwidth]{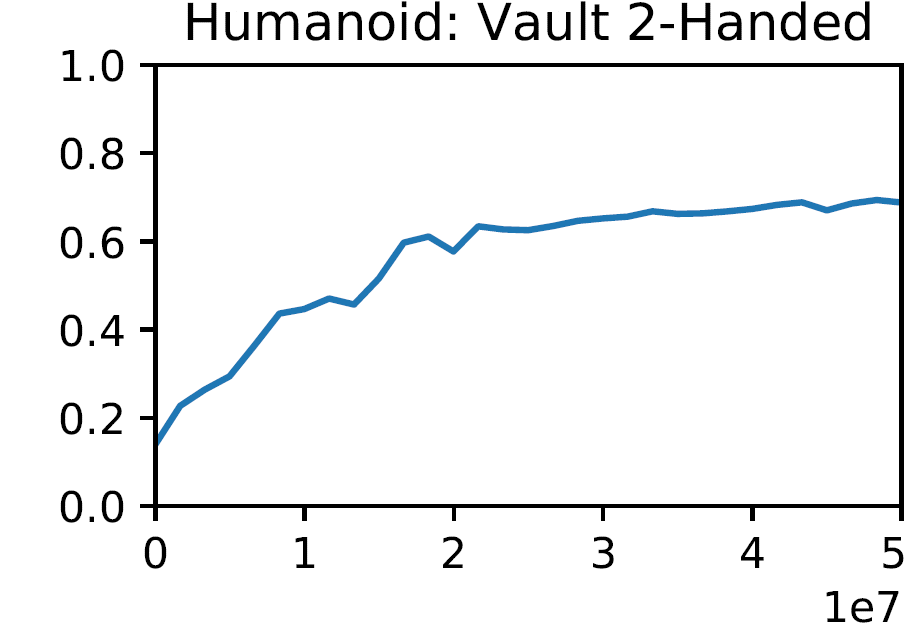}}
     	\subfigure{\includegraphics[width=0.5\columnwidth]{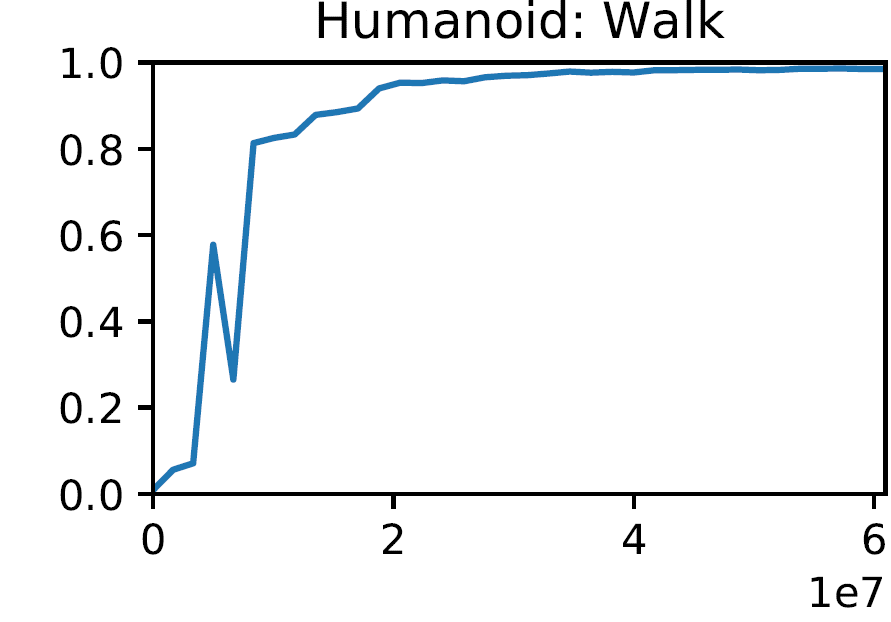}}
\caption{Learning curves of the humanoid imitating individual motion clips. Performance is calculated as the mean return over 32 episodes. The returns are normalized by the minimum and maximum possible return per episode. Due to the time needed to train each policy, performance statistics are collected only from one run of the training process for the majority of skills. For the backflip and run policies, we collected statistics from 3 training runs using different random seeds. Performance appears consistent across multiple runs, and we have observed similar behaviours for many of the other skills.}
\label{fig:allLearningCurves0}
\end{figure*}

\begin{figure*}[h]
	\flushleft
    	\subfigure{\includegraphics[width=0.5\columnwidth]{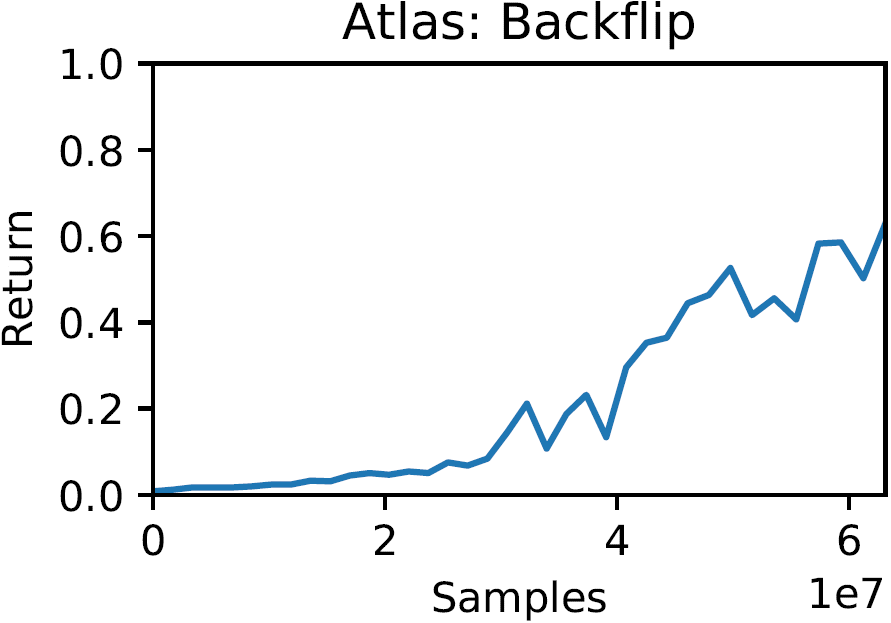}}
    	\subfigure{\includegraphics[width=0.5\columnwidth]{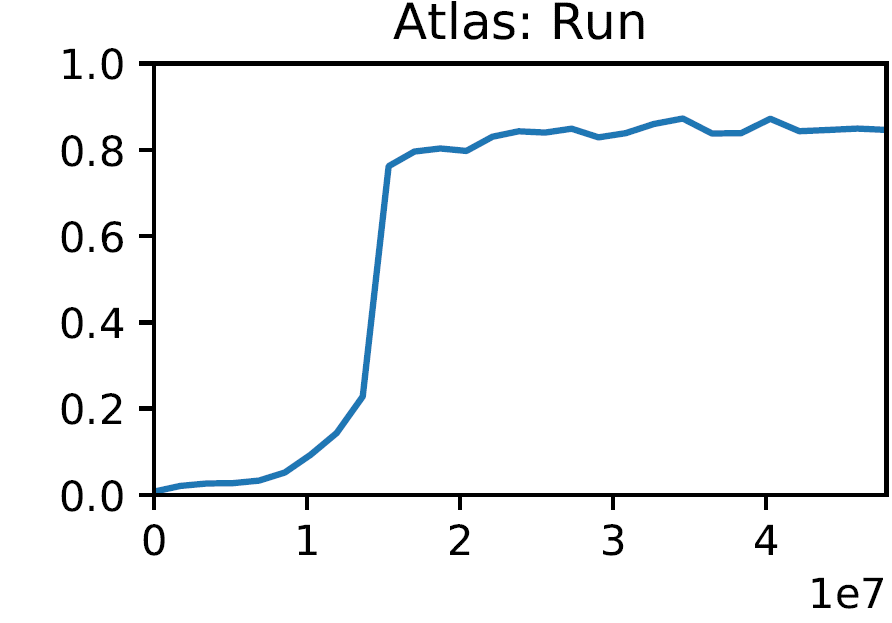}}
		\subfigure{\includegraphics[width=0.5\columnwidth]{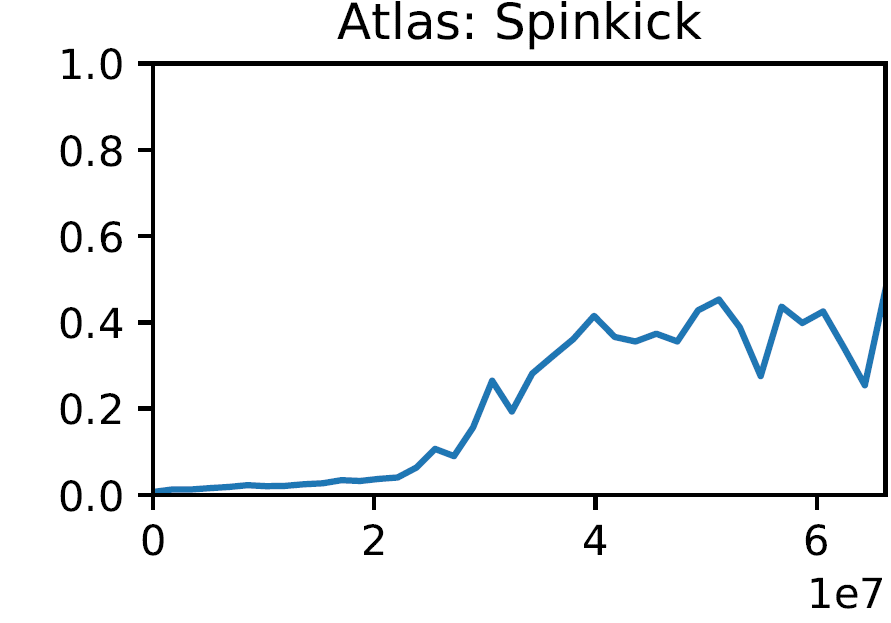}}
    	\subfigure{\includegraphics[width=0.5\columnwidth]{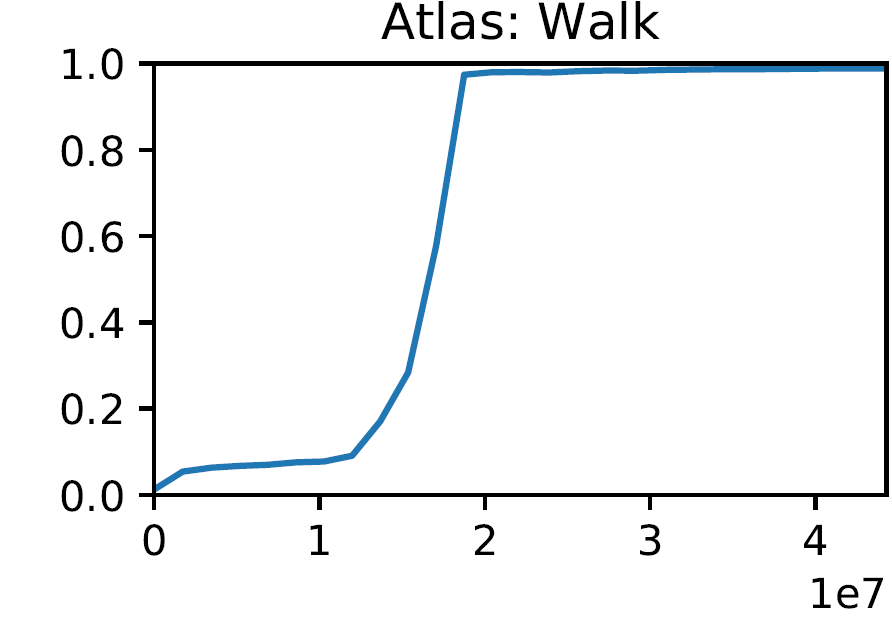}}
    \subfigure{\includegraphics[width=0.5\columnwidth]{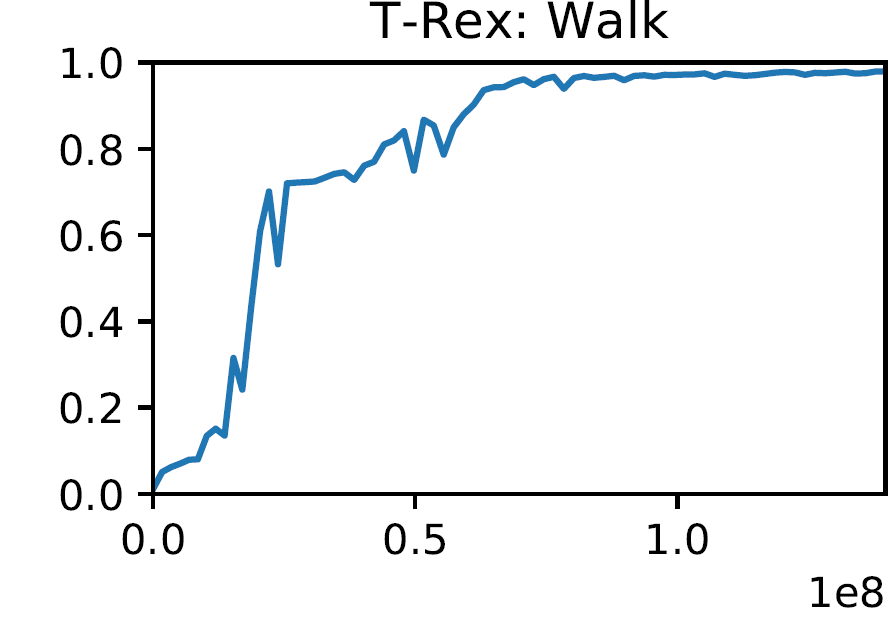}}
    \subfigure{\includegraphics[width=0.5\columnwidth]{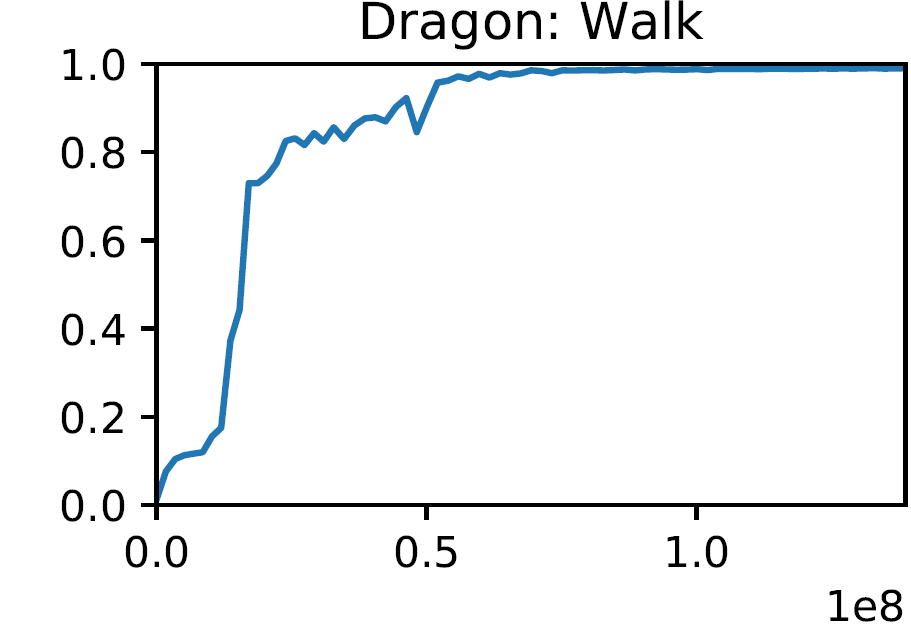}}
\caption{Learning curves of the Atlas, T-Rex, and dragon imitating individual motion clips.}
\label{fig:allLearningCurves1}
\end{figure*}

\begin{figure*}[h]
	\flushleft
    	\subfigure{\includegraphics[width=0.5\columnwidth]{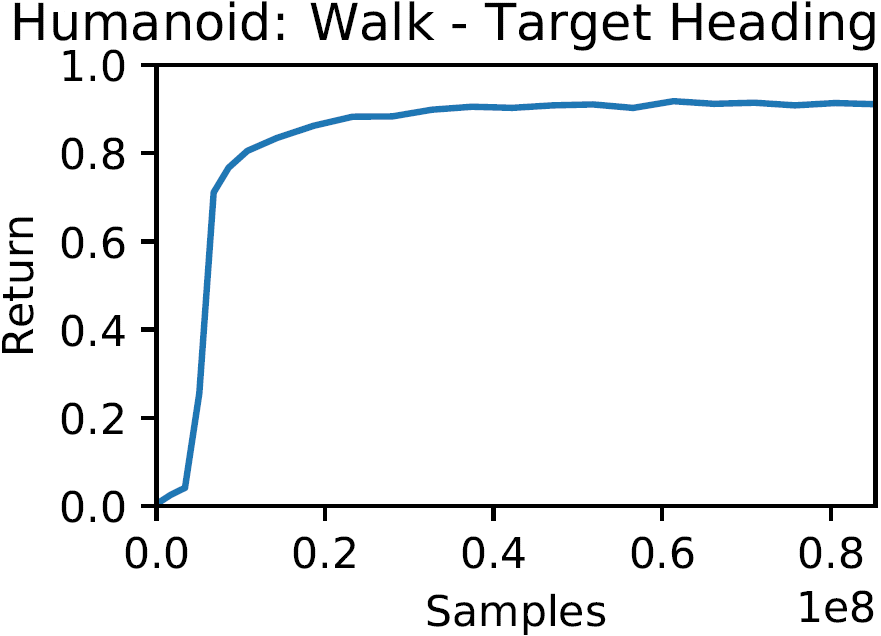}}
    	\subfigure{\includegraphics[width=0.5\columnwidth]{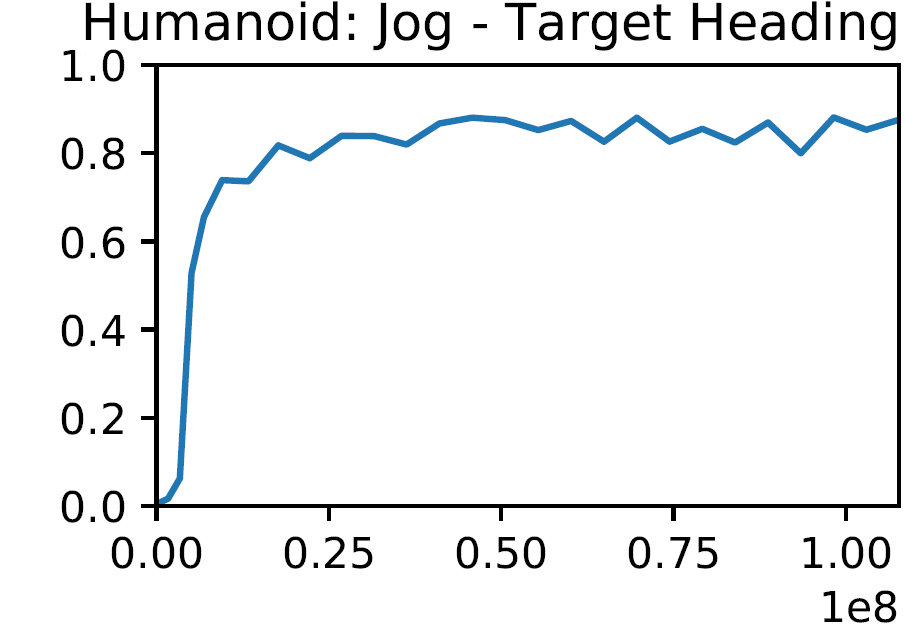}}
		\subfigure{\includegraphics[width=0.5\columnwidth]{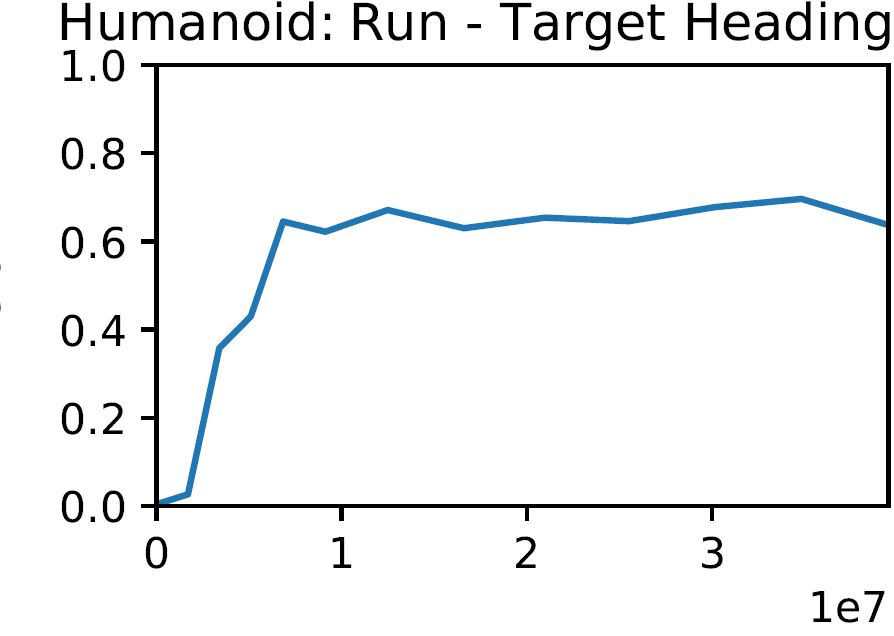}}
    	\subfigure{\includegraphics[width=0.5\columnwidth]{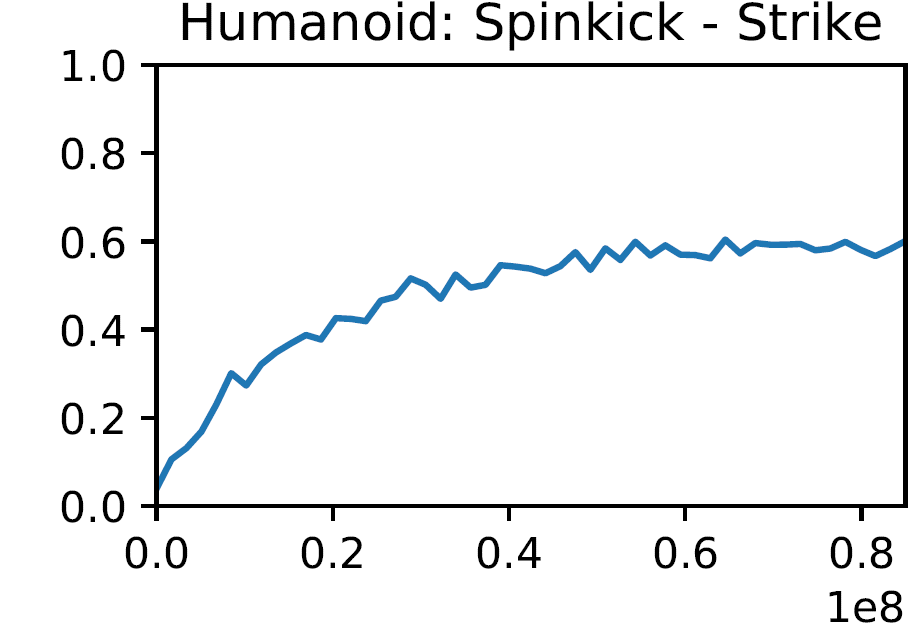}}
    \subfigure{\includegraphics[width=0.5\columnwidth]{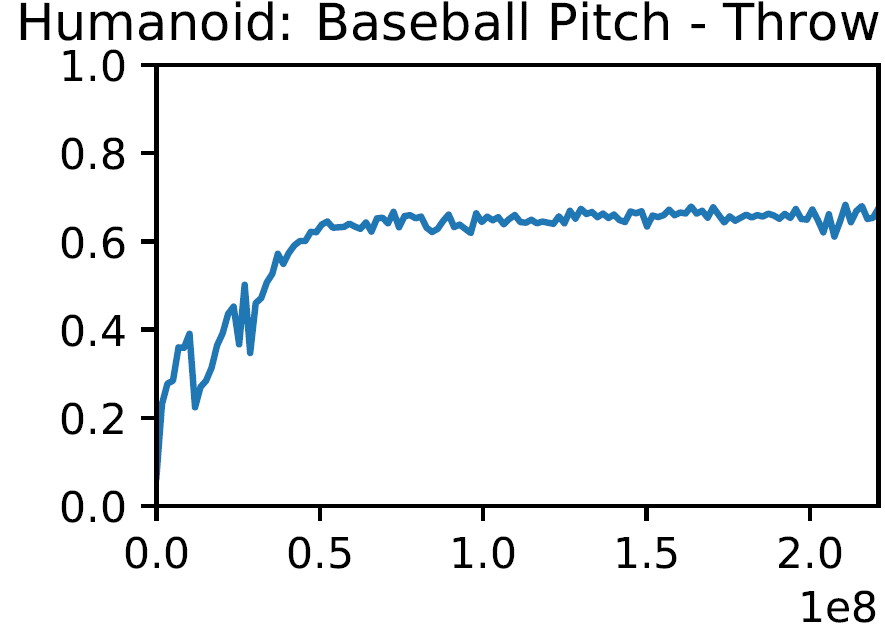}}
    \subfigure{\includegraphics[width=0.5\columnwidth]{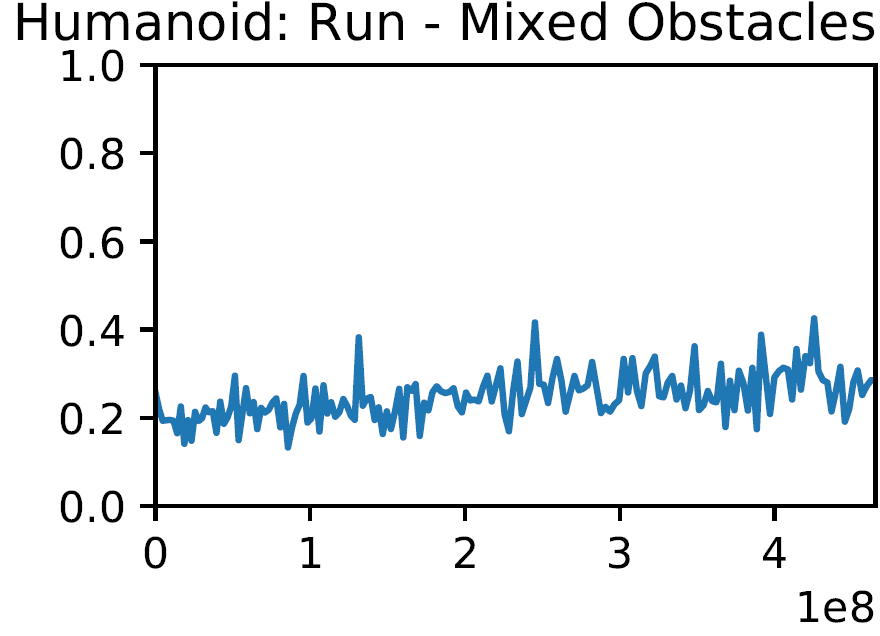}}
    \subfigure{\includegraphics[width=0.5\columnwidth]{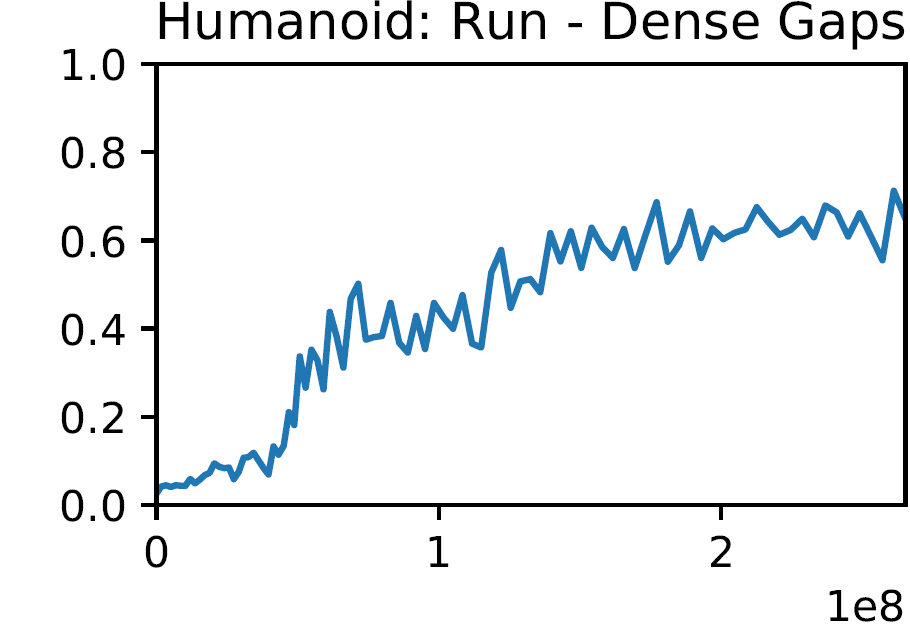}}
    \subfigure{\includegraphics[width=0.5\columnwidth]{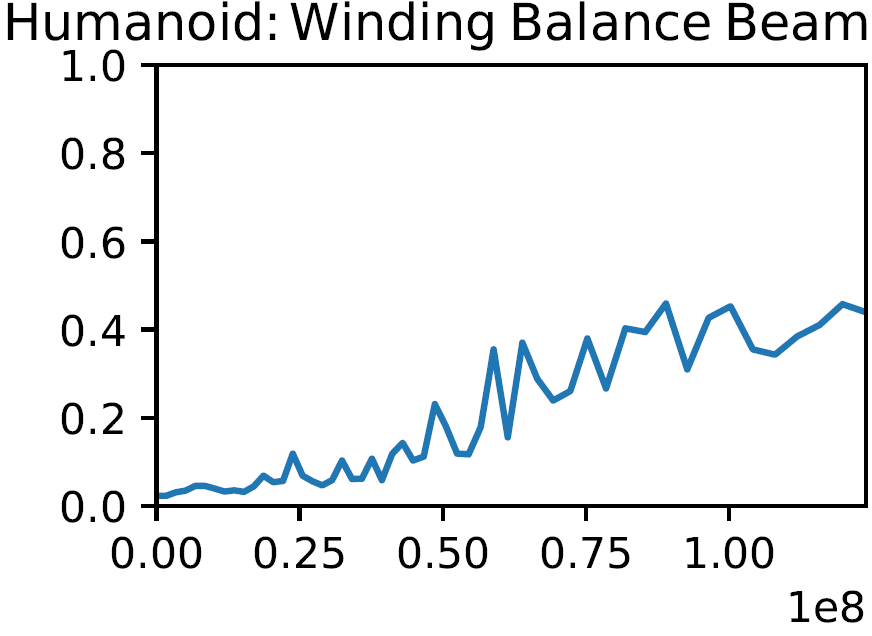}}
    \subfigure{\includegraphics[width=0.5\columnwidth]{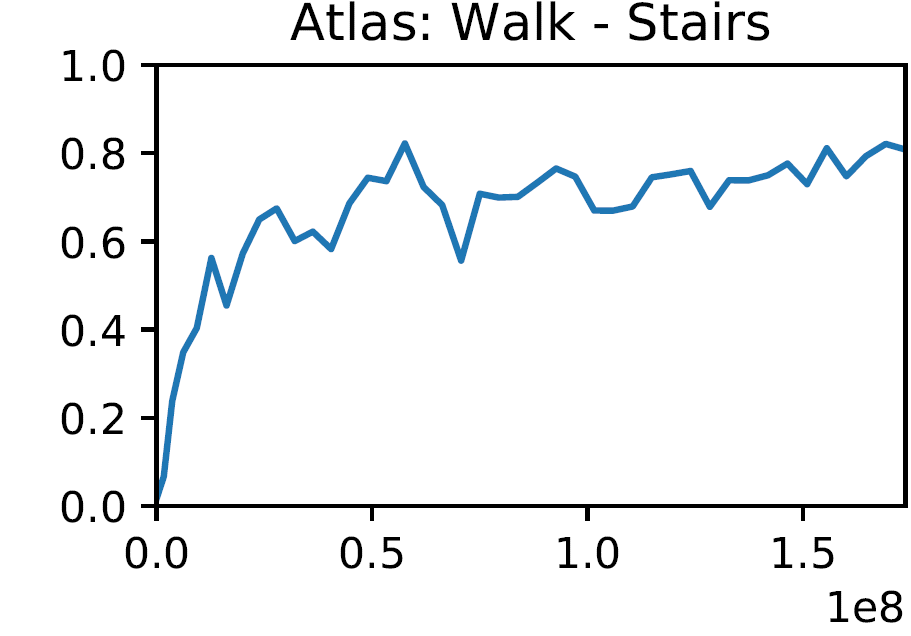}}
    \subfigure{\includegraphics[width=0.5\columnwidth]{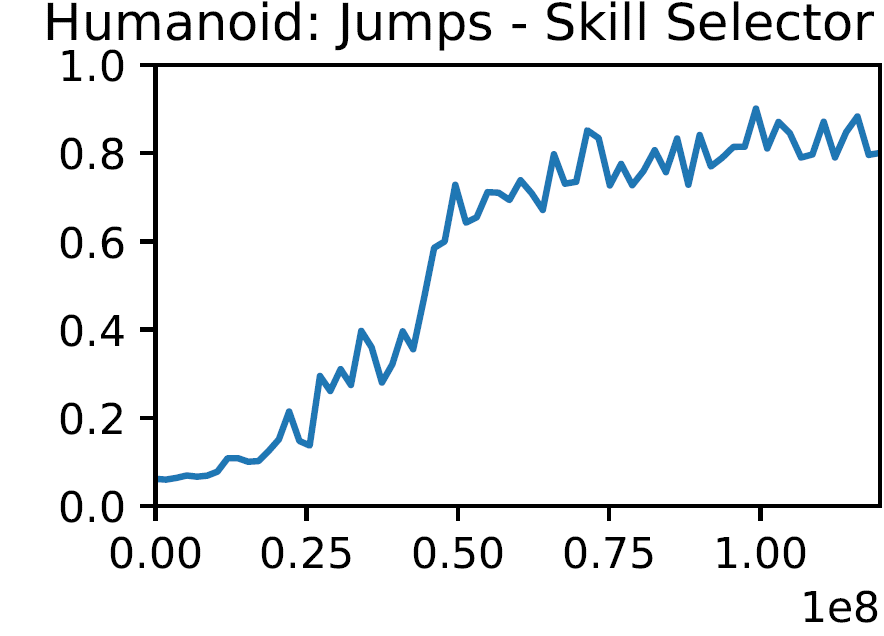}}
    \subfigure{\includegraphics[width=0.5\columnwidth]{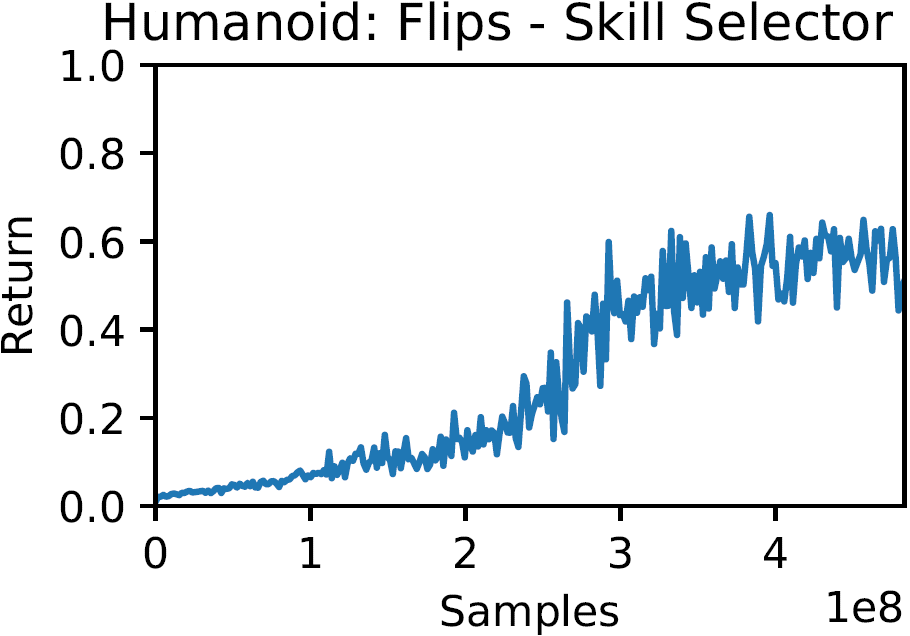}}
\caption{Learning curves of policies trained to imitate motion clips while fulfilling additional task objectives.}
\label{fig:allLearningCurves2}
\end{figure*}

\end{document}